\documentclass[aps,prl,english,superscriptaddress,twocolumn,tightenlines,showpacs,longbibliography,floatfix,amsfonts]{revtex4-2}
\usepackage[T1]{fontenc}
\usepackage[utf8]{inputenc}
\usepackage[normalem]{ulem}
\usepackage{textcomp}
\usepackage{graphicx,float}
\usepackage[dvipsnames,svgnames]{xcolor}
\usepackage{tikz}
\usetikzlibrary{arrows.meta}
\usepackage{mathrsfs,mathtools,xfrac}
\usepackage{amsmath,amssymb,amsthm,amscd,bm,bbm}
\usepackage{txfonts} 
\usepackage{afterpackage,multirow,comment}
\usepackage{array,booktabs,minibox}
\newcolumntype{P}[1]{>{\centering\arraybackslash}p{#1}}
\usepackage[colorlinks,bookmarks=false,citecolor=blue,linkcolor=blue,urlcolor=blue,filecolor=blue]{hyperref}
\usepackage{wasysym}

\newtheorem{prop}{Proposition}
\newtheorem{cor}{Corollary}

\newcommand{\im}[1]{{\operatorname{im} \, #1}}
\newcommand{\Ann}[1]{{\operatorname{Ann} \, #1}}

\DeclareMathOperator{\supp}{supp}

\let\SMfigurecaptionname\figurename
\let\SMtablecaptionname\tablename

\usepackage{babel}

\renewcommand{\im}[1]{{\operatorname{im} #1}}

\renewcommand{\Ann}[1]{{\operatorname{Ann} #1}}

\newcolumntype{C}{>{$}c<{$}}
\AtBeginDocument{
	\heavyrulewidth=.08em
	\lightrulewidth=.05em
	\cmidrulewidth=.03em
	\belowrulesep=.65ex
	\belowbottomsep=0pt
	\aboverulesep=.4ex
	\abovetopsep=0pt
	\cmidrulesep=\doublerulesep
	\cmidrulekern=.5em
	\defaultaddspace=.5em
}

\begin{document}

\title{Algebraic Framework and Optimal Thresholds for Topological Codes in Spacetime}

\author{Bai-Bo Wang}
\thanks{These authors contributed equally.}
\affiliation{Hefei National Research Center for Physical Sciences at the Microscale and School of Physical Sciences, University of Science and Technology of China, Hefei 230026, China}
\affiliation{Hefei National Laboratory, University of Science and Technology of China, Hefei, 230088, China}

\author{Ji-Ze Xu}
\thanks{These authors contributed equally.}
\affiliation{Hefei National Research Center for Physical Sciences at the Microscale and School of Physical Sciences, University of Science and Technology of China, Hefei 230026, China}
\affiliation{Hefei National Laboratory, University of Science and Technology of China, Hefei, 230088, China}

\author{Hai-Jun Liao}
\affiliation{Beijing National Laboratory for Condensed Matter Physics and Institute of Physics,
Chinese Academy of Sciences, Beijing 100190, China.}

\author{Hao Song}
\email{songhao@itp.ac.cn}
\affiliation{Institute of Theoretical Physics, Chinese Academy of Sciences, Beijing 100190, China}

\author{Ke Liu}
\email{ke.liu@ustc.edu.cn}
\affiliation{Hefei National Research Center for Physical Sciences at the Microscale and School of Physical Sciences, University of Science and Technology of China, Hefei 230026, China}
\affiliation{Shanghai Research Center for Quantum Science and CAS Center for Excellence in Quantum Information and Quantum Physics, University of Science and Technology of China, Shanghai 201315, China}

\date{\today}
	
\begin{abstract}
Quantum error-correcting codes offer increasingly versatile routes to fault-tolerant quantum computation, but their intrinsic noise resilience remains difficult to quantify. 
In this work, we develop an algebraic framework that unifies topological protection, optimal-threshold computation, and gauge-structure analysis in spacetime. Through a free-resolution construction of the spacetime fault complex and exactness transfer across CSS sectors, our framework establishes local stabilizer data as the complete algebraic input for determining optimal thresholds and code-induced gauge structures. Computational machinery built on this framework enables calculations in the thermodynamic limit for structurally complex codes.
The application to six prominent codes spanning qLDPC, higher-dimensional topological, and fracton families yields previously unknown optimal thresholds, establishing their intrinsic error resilience. 
Our results cover more recent bivariate bicycle code families and resolve long-standing threshold problems for the three-dimensional color code and Haah's code.
Moreover, we uncover new spacetime gauge theories with higher-form, translation-enriched, subsystem, and fractal structures and demonstrate a general construction of their order parameters from gauge-charge neutrality.
Our work provides a constructive foundation for assessing the fault-tolerance potential of quantum stabilizer codes and for the systematic discovery of unconventional gauge theories.
\end{abstract}
	
\maketitle

\emph{Introduction}---
Quantum error-correcting codes increasingly exploit intricate stabilizer structures and nonlocal connectivity to achieve more versatile logical operations and higher encoding efficiency~\cite{Campbell17,Bombin07,Haah11,Kovalev13,Tillich14,Bombin15,Vijay16,Krishna21,Breuckmann21b,Panteleev22b,Leverrier22,Bravyi24,Xu24,Williamson24}.
These advantages become consequential for fault-tolerant quantum computation only if the codes are sufficiently resilient to noise.
A code’s noise resilience depends on the collective behavior of an ensemble of error configurations, not merely on the minimum weight of a logical error.
The optimal threshold quantifies this resilience by marking the code's correctability boundary and governing logical-error suppression~\cite{Dennis02,Chubb21}.
Nevertheless, it is undetermined even for prominent code families.

Determining the optimal threshold is both theoretically revealing and technically challenging.
Optimal decoding admits a statistical-mechanical description in which the loss of correctability corresponds to a phase transition in a spin model with quenched disorder~\cite{Dennis02}.
This correspondence has long driven efforts to locate optimal thresholds and explore unconventional states of matter~\cite{Wang03,Katzgraber09,Andrist_2011,Bombin12,Kubica18,Song22,Canossa26}.
However, the standard route has relied heavily on model-specific analysis and large-scale numerical simulations.
For structurally complex codes, it can entail substantial analytical effort and prohibitive computational cost~\cite{Song22,Canossa26}.
Recent advances have broadened diagnostic approaches and representations of fault-tolerance properties in quantum codes~\cite{Fan24,Su24,Colmenarez24,Lyons24,Li25,Hillmann25,Aitchison25,Xu26b,Lee26}.
They remain largely focused on benchmarking against known thresholds and establishing code-distance scaling.
Yet frameworks that provide a general formulation of code fault tolerance and enable systematic quantification of error resilience are still elusive.

In this Letter, we develop a constructive algebraic framework for determining optimal thresholds of topological codes in spacetime.
Going beyond representation, the framework establishes topological protection in spacetime and unifies threshold computation and gauge-structure analysis.
This unification is achieved through a free-resolution construction of the spacetime fault complex and the transfer of exactness across CSS sectors to the dual interaction complex.
Within this framework, we construct computational machinery in which local stabilizer data supply the complete algebraic input for determining optimal thresholds, even for structurally complex codes.
From the same data, the framework also specifies the complete local gauge structure encoding the spacetime constraints on code excitations.

To demonstrate its reach, we compute the previously unknown optimal thresholds for six prominent codes under spacetime decoherence, summarized in Table~\ref{tab:thresholds}.
These codes span structurally distinct families and include more recent bivariate bicycle (BB) codes alongside computationally challenging examples such as the 3D color code and Haah's code, introduced in 2006 and 2011, respectively~\cite{Bombin07,Haah11}.
The two BB families containing the gross and two-gross codes both yield $p_{\mathrm{th}}=0.029(2)$, despite their different maximal logical-qubit counts.
For the 3D color code, a paradigmatic route to fault-tolerant non-Clifford gates, we obtain $p_{\mathrm{th}}=0.011(1)$, resolving its long-standing threshold problem.
Moreover, we reveal exceptional error resilience of fracton codes, with $p_{\mathrm{th}}=0.034(2)$, $0.034(1)$, and $0.022(1)$ for Haah's, checkerboard, and X-cube codes, respectively.

In addition, the codes investigated here induce new spacetime gauge theories with higher-form, symmetry-enriched, subsystem, and fractal structures.
We construct their gauge-invariant order parameters and demonstrate a fundamental distinction between gauge-charge closure and geometric closure.

\begin{table}[t]
\caption{Optimal thresholds and induced gauge theories for six representative codes under spacetime decoherence, spanning qLDPC, higher-dimensional topological, and fracton families.
All these thresholds were previously unknown and would be difficult to determine using conventional approaches.
The code-induced gauge theories exhibit novel gauge structures and order parameters.}
\label{tab:thresholds}
\centering
\footnotesize
\setlength{\tabcolsep}{0pt}
\renewcommand{\arraystretch}{1.618}
\newcommand{\thresholdtablehead}[1]{%
	\begingroup
	\bfseries
	\renewcommand{\arraystretch}{1}%
	\begin{tabular}[c]{@{}c@{}}#1\end{tabular}%
	\endgroup}
\newcommand{\thresholdcodecell}[1]{\makebox[0.34\columnwidth][c]{#1}}
\newcommand{\thresholdvaluecell}[1]{\makebox[0.28\columnwidth][c]{#1}}
\newcommand{\thresholdgaugecell}[1]{\makebox[0.38\columnwidth][c]{#1}}
\begin{tabular}{ccc}
\toprule
\thresholdcodecell{\thresholdtablehead{Code}}
& \thresholdvaluecell{\thresholdtablehead{Optimal\\Threshold}}
& \thresholdgaugecell{\thresholdtablehead{Induced\\Gauge Theory}} \\
\colrule
\thresholdcodecell{$\mathbb{BB}(\bar 1,\bar 1,3,3)$}
& \thresholdvaluecell{$0.029(2)$}
& \multirow[c]{2}{*}{\thresholdgaugecell{\shortstack{3D, symmetry\\enriched}}} \\
\thresholdcodecell{$\mathbb{BB}(\bar 1,\bar 1,3,\bar 3)$}
& \thresholdvaluecell{$0.029(2)$} & \\
\thresholdcodecell{3D color} & \thresholdvaluecell{$0.011(1)$}
& \thresholdgaugecell{4D, higher-form} \\
\thresholdcodecell{Haah} & \thresholdvaluecell{$0.034(2)$}
& \thresholdgaugecell{4D, fractal} \\
\thresholdcodecell{Checkerboard} & \thresholdvaluecell{$0.034(1)$}
& \thresholdgaugecell{4D, subsystem} \\
\thresholdcodecell{X-cube} & \thresholdvaluecell{$0.022(1)$}
& \thresholdgaugecell{4D, subsystem} \\
\bottomrule
\end{tabular}
\end{table}

\emph{Algebraic spacetime framework}---
We first recall the standard algebraic description of a
translation-invariant Calderbank-Shor-Steane (CSS) code~\cite{Haah13}.
Let $R_{\rm s}=\mathbb{F}_2[x_1^{\pm1},\ldots,x_D^{\pm1}]$ be the Laurent polynomial ring of translations on a $D$-dimensional spatial lattice, and suppose each unit cell contains $a$ $X$-type checks, $b$ $Z$-type checks, and $q$ qubits.
The static code can be represented by the chain complex
\begin{align}
    C_1
    \xrightarrow{\partial_1}
    C_0
    \xrightarrow{\partial_0}
    C_{-1}.
    \label{eq:static_map}
\end{align}
Here $C_1=R_{\rm s}^a\oplus R_{\rm s}^b$, $C_0=R_{\rm s}^q\oplus R_{\rm s}^q$, and
$C_{-1}=R_{\rm s}^b\oplus R_{\rm s}^a$ are the free $R_{\rm s}$-modules of check labels,
Pauli supports, and syndromes, respectively.  
The differentials
$\partial_1=\begin{psmallmatrix}A&0\\0&B\end{psmallmatrix}$
and
$\partial_0=\begin{psmallmatrix}B^\dagger&0\\0&A^\dagger\end{psmallmatrix}$
are the stabilizer and syndrome maps, respectively, where $A$ and $B$
specify the supports of the $X$- and $Z$-type stabilizers.
The code is topological if the complex is exact on the infinite
lattice, namely, $\operatorname{im}\partial_1=\ker\partial_0$.

Promoting the chain complex to spacetime is necessary because realistic syndrome measurements are intrinsically noisy.
To obtain a unified description of the stabilizer structure and repeated syndrome measurements, we construct a spacetime fault complex:
\begin{align}\label{eq:spacetime_complex}
	F_\bullet:\quad
     \cdots\longrightarrow F_2
    \xrightarrow{\Square_2}F_1
    \xrightarrow{\Square_1}F_0
    \xrightarrow{\Square_0}F_{-1}\to0.
\end{align}
Here, the spacetime modules
$F_i=(C_i\oplus C_{i-1})\otimes_{R_{\rm s}}R_{\rm st}$
combine two adjacent modules of the static complex, with spacetime ring 
$R_{\rm st}=\mathbb F_2[x_1^{\pm1},\ldots,x_D^{\pm1},t^{\pm1}]$ and time translation $t$.
Two successive measurement rounds are coupled through the spacetime differentials 
\begin{align}\label{eq:spacetime_maps}
    \Square_{i\geq 1}=\begin{pmatrix}
        \partial_i & (t-1)I\\
        0 & \partial_{i-1}
    	\end{pmatrix},     \quad
    \Square_{0}=\begin{pmatrix}
        \partial_0 & (t-1)I
		\end{pmatrix}. 
\end{align}
The physical interpretation of this construction is transparent: $F_0$ is
the fault-history module, containing both data-qubit and measurement faults,
while $F_1$ and $F_{-1}$ label local fault deformations and detection events,
respectively.
The differentials $\Square_1$ and $\Square_0$ are the fault-deformation and detector maps, respectively.
The higher differentials $\Square_{i\geq 2}$ generate a hierarchy of local relations and complete a free resolution of the fault-deformation module.
The static complex is correspondingly extended to
$\thickmuskip=2mu C_\bullet:\cdots\to C_2\xrightarrow{\partial_2}C_1 \xrightarrow{\partial_1} C_0 \xrightarrow{\partial_0} C_{-1}$.

This formulation also makes the spacetime exactness relations and their
physical consequences transparent.
The spacetime fault complex in Eq.~\eqref{eq:spacetime_complex} factors as
$F_\bullet=C_\bullet\otimes_{R_{\rm s}}K_\bullet$, where repeated syndrome measurements are encoded by a single-generator Koszul complex $K_\bullet=[R_{\rm st}\xrightarrow{t-1}R_{\rm st}]$.
As $K_\bullet$ resolves $R_{\rm st}/(t-1)\cong R_{\rm s}$, the temporal
extension preserves static homology: $H_i(F_\bullet)\cong H_i(C_\bullet)$.
The static topological condition thus extends to spacetime as
$\operatorname{im}\Square_1=\ker\Square_0$, ensuring topological protection
and a diverging fault distance as the spacetime system grows.
Likewise, the higher exactness of $C_\bullet$ transfers to
$\operatorname{im}\Square_{i+1}=\ker\Square_i$ for $i\ge1$, ensuring that
the free resolution of $\operatorname{im}\Square_1$ encodes complete information about local fault deformations and their higher relations.
See SM~\cite{SM} for proofs.

Whereas $F_\bullet$ describes the spacetime stabilizer structure, the optimal
decoding problem is encoded in a dual formulation.
To faithfully represent fault-class likelihoods, the formulation must
account for the full kernel of $\Square_1^\dagger$ and its higher relations.
However, the ordinary dual of $F_\bullet$ does not generally inherit its exactness.
We need to supply the missing generators to remove the cohomological obstruction to completeness.

The resulting formulation is an interaction cochain complex.
For clarity, we decompose the CSS fault complex sectorwise as
$F_\bullet=F_\bullet^X\oplus F_\bullet^Z$ by taking the two diagonal blocks
of $\partial_1$ and $\partial_0$.
The most relevant
segment is
\begin{equation}\label{eq:paired_interaction_segment}
 \widetilde{F}_2^\nu
 \xrightarrow{\widetilde{\Square}_2^\nu}
\ F_0^\mu
 \xrightarrow{\widetilde{\Square}_1^\nu = (\Square_1^\mu)^\dagger}
 F_1^\mu
 \xrightarrow{(\Square_2^\mu)^\dagger}
 F_2^\mu,
\end{equation}
with $\{\mu,\nu\}=\{X,Z\}$ and 
$\widetilde F_1^\nu = (C_{0}^\nu\oplus C_1^\nu)\otimes_{R_{\rm s}}R_{\rm st} \cong F_0^\mu$ 
written in the reversed block order.
On this adjoint side, $(\Square_1^\mu)^\dagger$ defines the interaction map of the auxiliary-spin model underlying optimal decoding, while $(\Square_2^\mu)^\dagger$ encodes its gauge structure.
To generate all finite-support interaction relations, we construct
$\widetilde{\Square}_2^\nu=
\begin{psmallmatrix}
\partial_1^\nu & 0\\
(t^{-1}-1)I & \partial_2^\nu
\end{psmallmatrix}
= \begin{psmallmatrix}
(\Square_0^\mu)^\dagger & \big| &
\begin{smallmatrix}
0\\
\partial_2^\nu
\end{smallmatrix}
\end{psmallmatrix}$
 as a complete syzygy map of
$(\Square_1^\mu)^\dagger$, augmenting the detector adjoint
$(\Square_0^\mu)^\dagger$ with the missing generators.

This dual construction reveals a cross-sector transfer of completeness.
Exactness transfers from the $\nu$-sector fault complex to the complementary $\mu$-sector interaction cochain complex:
$\operatorname{im}\widetilde{\Square}_2^\nu\!=\ker(\Square_1^\mu)^\dagger$.
Consequently, the completion overcomes the ordinary adjoint obstruction,
$H^0((F_\bullet^\mu)^\dagger)\cong H^{-1}((C_\bullet^\mu)^\dagger)\cong\ker\partial_1^\nu$,
establishing exactness at $F_0^\mu$.
This correspondence extends to higher degrees through
$\widetilde{\Square}_i^\nu=\begin{psmallmatrix}
\partial_{i-1}^\nu&0\\
(t^{-1}-1)I&\partial_i^\nu
\end{psmallmatrix}$, $i\ge1$, obtained from $\Square_i^\nu$ by temporal
inversion and block exchange.
The temporal Koszul extension accounts for the one-degree shift in adjoint cohomology,
$H^j((F_\bullet^\mu)^\dagger)\cong H^{j-1}((C_\bullet^\mu)^\dagger)$.
CSS pairing and exactness transport thus supply the complete interaction-relation hierarchy without a separate construction.
In particular, $H^1((F_\bullet^\mu)^\dagger)\cong H_0(C_\bullet^\nu)=0$ ensures the same local data also generate all finite-support gauge-invariant observables~\cite{SM}.

The framework thus unifies spacetime fault tolerance, threshold computation, and gauge-structure analysis in a constructive algebraic formulation.

\emph{Computational machinery for optimal thresholds}---
To demonstrate the framework's computational capability beyond formal representation, we develop machinery for threshold determination from the spacetime boundary maps.

Let $\xi\in F_0$ be a fault history and $[\xi]=\xi+\operatorname{im}\Square_1$ denote its fault-equivalence class.
Optimal decoding compares the probabilities of classes compatible with the observed detection-event record.
For clarity, we treat the two CSS sectors separately and assume independent qubit and measurement faults with a common error rate $p$.
We assign each fault-equivalence class $[\xi_\mu]$ a partition function, defined by
\begin{align}\label{eq:Z}
    \mathcal Z_{\xi_\mu}(\beta)
    &=\sum_{\zeta_\mu}
    \exp\!\left[
    \beta\sum_\ell
    (-1)^{\xi_\mu(\ell)+(\Square_1^\mu\zeta_\mu)(\ell)}
    \right] \nonumber\\
    &=Z_0(\beta)
    \sum_{g\in\ker((\Square_1^\mu)^\dagger)}
    (\tanh\beta)^{|g|}
    \prod_{\ell\in\operatorname{supp}g}J_\ell .
\end{align}
Here, $\ell$ labels fault locations in spacetime, and $\zeta_\mu\in F_1^\mu$ parametrizes local fault deformations.
The parameter $\beta$ is an effective inverse temperature defined by the Nishimori condition $e^{-2\beta}=\frac{p}{1-p}$, and $J_{\ell} = (-1)^{\xi_\mu(\ell)} = \pm 1$ denotes the quenched random coupling at location $\ell$ in the underlying spin model, whereas the interaction pattern is specified by $(\Square_1^\mu)^\dagger$.
$Z_0(\beta)$ is a nonsingular background.

This formulation connects the correctability of a code to a boundary-generated graph expansion.
Admissible graphs $g$ are inherited from the interaction cochain complex through the kernel condition $(\Square_1^\mu)^\dagger g=0$.
Exactness at $F_0^\mu$ in Eq.~\eqref{eq:paired_interaction_segment} then ensures a complete, constructive parametrization $g=\widetilde{\Square}_2^\nu\gamma$ in the thermodymial limit, with $\gamma$ a finite configuration of local generators supplied by the stabilizer structure.
The graph-generation procedure is entirely algebraic, applies across distinct codes without requiring code-specific geometric constructions or tessellations, and is independent of the disorder realization.

The optimal threshold corresponds to a thermodynamic phase transition on the Nishimori line, signaled by a nonanalyticity in the quenched free energy.
We compute the quenched free-energy density $\big[f_{\xi_\mu}\big]$ by integrating M\"obius inversion over generator subsets $\gamma$ with analytic disorder averaging, yielding
\begin{align} \label{eq:free-energy}
	\beta \big[f_{\xi_\mu}\big] =  \beta f_0 - \sum_m \sum_{k} A_{m,k}(1-2p)^k\tanh^m \beta .
\end{align}
Here, $f_0$ is the nonsingular contribution from $Z_0(\beta)$, $m$ is the expansion order, $k\in\{m,m-2,\ldots,m\bmod 2\}$ arises from analytic disorder averaging, and $A_{m,k}$ is the corresponding expansion coefficient.
These coefficients are computed directly in the thermodynamic limit, so no finite-size extrapolation is required.
Moreover, retaining $p$ symbolically makes the expansion reusable across disorder strengths, and analytic disorder averaging eliminates the need to sample individual realizations. See SM~\cite{SM} for details.

This algebraic machinery overcomes key bottlenecks that conventional approaches face in model construction and computation, especially for structurally complex codes.

\begin{figure}[t]
    \centering
    \includegraphics[width=0.98\columnwidth]{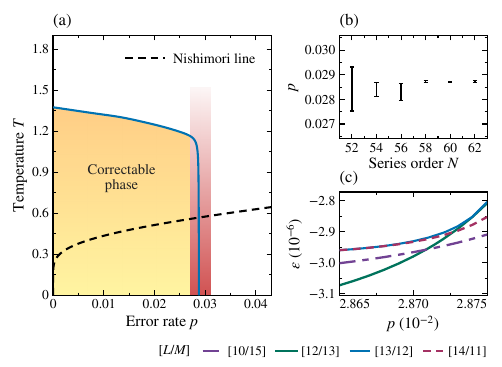}
    \caption{Optimal threshold of the
    $\mathbb{BB}(\bar 1,\bar 1,3,3)$ code under spacetime decoherence.
    (a) The phase boundary is determined by Dlog--Pad\'e analysis of
    the graph expansion through order $N=62$.
    Its intersection with the Nishimori line (black dashed) locates
    the optimal threshold. The optimal threshold is
    $p_{\mathrm{th}}=0.029(2)$, with its uncertainty range shown by
    the red shading.
    (b) Order dependence of the threshold estimates across expansion
    orders $N=52,54,\ldots,62$, using 26 different Dlog--Pad\'e
    approximants in total. At each order, the black bar spans the
    minimum and maximum estimates across the approximants.
    (c) Reconstruction of the next coefficient from lower-order series
    data through $N=60$. The small deviation $\varepsilon$ from the
    coefficient directly computed at $N=62$ provides a consistency
    check on the extrapolation; see SM~\cite{SM} for details.}
    \label{fig:bb1133-maintext}
\end{figure}

\emph{Optimal thresholds for representative codes}---
We compute previously unknown optimal thresholds for six codes spanning qLDPC, higher-dimensional topological, and fracton families; the results are summarized in Table~\ref{tab:thresholds}.
The overall threshold of each code is set by the minimum threshold of the two CSS sectors, $p_{\mathrm{th}}=\min\!\left(p_{\mathrm{th}}^{X},p_{\mathrm{th}}^{Z}\right)$.
Details of the calculation and threshold extraction are provided in the SM~\cite{SM}.

For benchmarking, we also reproduce the code-capacity thresholds of the 3D color and X-cube codes in the perfect-measurement limit.
We obtain threshold values of $0.020(2)$ and $0.074(2)$, respectively, in excellent agreement with the parallel-tempering Monte Carlo estimates of $0.019$ and $0.075(2)$~\cite{SM,Kubica18,Song22}.

Figure~\ref{fig:bb1133-maintext} illustrates threshold extraction for the BB code family $\mathbb{BB}(\bar1,\bar1,3,3)$.
The gross $[[144,12,12]]$ and two-gross $[[288,12,18]]$ codes featured in IBM's fault-tolerant roadmap are specific finite-size realizations of this family and the $\mathbb{BB}(\bar1,\bar1,3,\bar3)$ family, respectively~\cite{Bravyi24,Chen25,Yoder25Tour}.
Despite extensive searches for BB codes with favorable encoding rates and code distances, their optimal thresholds had remained undetermined.
Our machinery gives $p_{\mathrm{th}}=0.029(2)$ for both families, despite their notably different maximal logical-qubit counts, $k_{\max}=16$ and $26$, respectively~\cite{Chen25}.
The shared threshold estimate shows that greater maximal logical capacity can coexist with comparable intrinsic noise tolerance.

Another important example in Table~\ref{tab:thresholds} is the 3D color code, which represents a paradigmatic construction for realizing fault-tolerant non-Clifford logical gates~\cite{Bombin07}.
However, its intricate stabilizer structure and decoding difficulty have hindered determination of its optimal threshold beyond the code-capacity setting.
Whether its threshold is too low for practical hardware is a critical concern.
We obtain an optimal threshold $p_{\mathrm{th}}=0.011(1)$, establishing its intrinsic noise tolerance.

Among fracton codes, Haah's type-II code poses a particular challenge for threshold determination.
Its fractal symmetry leads to exponentially spaced finite-size sequences, such as $L=2^n$~\cite{Haah13,Canossa26}.
Our algebraic machinery overcomes this finite-size barrier and yields $p_{\mathrm{th}}=0.034(2)$ under spacetime decoherence.
For the type-I checkerboard and X-cube codes, we obtain optimal thresholds $p_{\mathrm{th}}=0.034(1)$ and $0.022(1)$, respectively.
The X-cube threshold is close to the 3D toric-code benchmark $p_{\mathrm{th}}\approx0.02$~\cite{Xu26}, while Haah's and checkerboard codes exhibit substantially higher thresholds.

\begin{figure}[t]
    \centering
    \includegraphics[width=0.98\columnwidth]{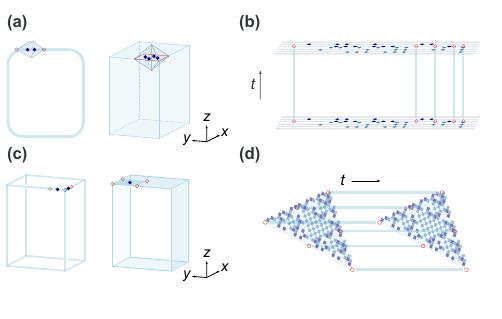}
    \caption{Gauge-invaraint order parameters for code-induced spacetime gauge theories.
	(a) 3D color code: a Wilson loop in the $X$ sector (left) and a closed membrane in the $Z$ sector (right). The tetrahedra illustrate local gauge-charge motifs.
	(b) BB code: Spatial ribbons of bounded width joined by temporal strands, illustrated for $\mathbb{BB}(\bar 1,\bar 1,3,3)$. The microscopic pattern within each ribbon reflects quasi-fractonic charge transport.
	(c) X-cube code: a wireframe-like support in the $Z$ sector (left) and an open-ended tube in the $X$ sector (right).
	(d) Haah’s code: two spatial fractal patterns joined by temporal strands, forming a sparse, non-manifold support.
	Spins are represented by diamonds, with different colors distinguishing their types. Dashed circles and polygons indicate charge excitations.}
    \label{fig:ops}
\end{figure}

\emph{Gauge structures and order parameters}---
The optimal-decoding models for the codes in Table~\ref{tab:thresholds} realize several new gauge theories with higher-form, translation-enriched, subsystem, and fractal structures.
These theories capture the spacetime constraints on code excitations and are of fundamental interest to many-body physics.
More generally, our framework determines the complete local gauge structure of code-induced theories and makes the construction of their gauge-invariant order parameters transparent.

The construction is entirely algebraic and requires only local stabilizer data.
Let $\Sigma\in F_1^\mu$ denote a finite spin support in the optimal-decoding model and $W_\Sigma$ the associated spin product.
The chain condition $\Square_1^\mu\Square_2^\mu=0$ ensures that the local transformations generated by $\Square_2^\mu$ preserve every interaction term, while the exactness relation $\operatorname{im}\Square_2^\mu=\ker\Square_1^\mu$ establishes the completeness of these local gauge transformations.
The adjoint map assigns the gauge-charge pattern $(\Square_2^\mu)^\dagger\Sigma$, whose nonzero entries identify the gauge charges carried by $W_\Sigma$.
Gauge invariance is therefore equivalent to charge neutrality, $(\Square_2^\mu)^\dagger\Sigma=0$, and the adjoint exactness relation $\operatorname{im}(\Square_1^\mu)^\dagger=\ker(\Square_2^\mu)^\dagger$ guarantees that elementary interactions generate all finite-support gauge-invariant observables.

An order parameter for the gauge theory must be gauge invariant and extend to arbitrary scales to probe long-distance charge behavior.
Our framework constructs gauge-invariant order parameters by combining charge creation, propagation, and annihilation in spacetime, subject to overall charge neutrality.
Given $F_\bullet=C_\bullet\otimes_{R_{\mathrm s}}K_\bullet$, the order-parameter support can be decomposed into spatial and temporal components whose gauge charges mutually cancel.
A convenient choice is to first solve for an extended spatial support $w$, then carry its excitation pattern to an arbitrary measurement round $r\geq1$.
The resulting spin support is

\begin{equation}\label{eq:order-parameter}
 \medmuskip=1mu \thickmuskip=2mu
 \Sigma^\mu_{(w;r)}=
 (\Square_1^\mu)^\dagger
 \begin{pmatrix}
   (t+\cdots+t^r)w\\
   0
 \end{pmatrix}
 =
 \begin{pmatrix}
   (t+\cdots+t^r)(\partial_1^\mu)^\dagger w\\
   (1+t^r)w
 \end{pmatrix}.
\end{equation}
Here, the upper and lower components describe the temporal propagation and spatial creation and annihilation of the static excitation pattern $(\partial_1^\mu)^\dagger w$, respectively.
The propagation joins successive rounds by canceling their paired creation and annihilation terms, forming a gauge-invariant history rather than merely copying the pattern $r$ times.

This charge-neutralization procedure is code-independent and does not require prescribing observable geometries or gauge-charge types.
Figure~\ref{fig:ops} illustrates characteristic order parameters for the codes in Table~\ref{tab:thresholds}, with construction details given in the SM~\cite{SM}.
The 3D color code leads to more familiar order parameters in 4D spacetime: Wilson loops in its $X$ sector and closed membranes in its $Z$ sector, reflecting the latter's two-form gauge structure.
BB codes instead induce translation-enriched gauge theories with ribbon order parameters. Although effectively one-dimensional at large scales, these ribbons have a finite width and a microscopic structure that reflects the quasi-fractonic anyon hopping of BB codes.

The order parameters of fracton-code-induced gauge theories reveal a fundamental distinction between charge closure and geometric closure.
The $Z$ sector of the X-cube code yields wireframe-like order parameters, while its $X$ sector and the checkerboard code yield geometrically open-ended tubes.
Their corresponding gauge charges exhibit restricted spatial mobility, characterizing distinct 4D subsystem gauge theories.
For Haah's code, the induced 4D gauge theory has a fractal gauge structure that encodes the code's spatial charge-creation rules and their temporal extension.
The associated order parameter has a sparse, non-manifold geometry, with two spatial fractal patterns joined by temporal strands.

\emph{Summary and outlook}---
In this Letter, we established a constructive algebraic framework for assessing the intrinsic error resilience of topological CSS codes and characterizing the gauge theories they induce. 
Our framework connects the fault and interaction complexes through complete relation hierarchies generated from the same local stabilizer data, enabling systematic threshold computation and gauge structure analysis.
We demonstrate this capability by determining previously unknown optimal thresholds and associated gauge structures for six prominent, structurally distinct quantum codes under spacetime decoherence. 
Each code presents a demanding problem in its own right, yet all six admit a common, direct treatment within our framework.

The framework can also be applied to a wider range of translation-invariant CSS codes that admit a topological description~\cite{Chen25,Yin25,Roeck25,Zhu25,Zhang26}.
This allows error resilience to be a parallel criterion alongside non-Clifford gate capabilities and high encoding rates.
In addition, recent work has developed statistical-mechanical mappings for noisy stabilizer circuits, with applications to the repetition and toric codes~\cite{Vodola22,Aitchison25}.
Extending our algebraic framework to encompass these circuit-level representations could open a route to assessing error resilience across different circuit implementations of structurally complex codes.

\begin{acknowledgments}
\emph{Acknowledgments}.
This work is supported by 
the Strategic Priority Research Program of Chinese Academy of Sciences (Grant No.~XDB1680000)
and the National Natural Science Foundation of China (Grants No.~12675032, 12522502, 12474145, 12447101, 12322403, 12347107), the CAS Project for Young Scientists in Basic Research (Grant No.~YSBR-150), the National Key Research and Development Project of China (Grants No.~2024YFA1408604 and 2022YFA1403900).

\end{acknowledgments}

\emph{Data availability}---
The series expansion coefficients and graph enumeration counts
are available at Zenodo~\cite{Wang2026Zenodo}, together with
a description of the method for determining sufficient storage cutoffs.

\bibliography{qec}
	

	\clearpage
	
	\onecolumngrid

	\begin{center}
		\textbf{\large --- Supplemental Material ---\\[0.5em]
			Algebraic Framework and Optimal Thresholds for Topological Codes in Spacetime
		}\\[1em]

		Bai-Bo Wang$^{1,2}$, Ji-Ze Xu$^{1,2}$, Hai-Jun Liao$^{3}$,
		Hao Song$^{4}$ and Ke Liu$^{1,5}$

		\vspace{5pt}

		{\centering \emph{$^1$Hefei National Research Center for Physical Sciences at the Microscale and School of Physical Sciences,\\
		University of Science and Technology of China, Hefei 230026, China}\par}

		{\centering \emph{$^2$Hefei National Laboratory, University of Science and Technology of China, Hefei, 230088, China}\par}

		{\centering \emph{$^3$Beijing National Laboratory for Condensed Matter Physics and Institute of Physics,\\
		Chinese Academy of Sciences, Beijing 100190, China.}\par}

		{\centering \emph{$^4$Institute of Theoretical Physics, Chinese Academy of Sciences, Beijing 100190, China}\par}

		{\centering \emph{$^5$Shanghai Research Center for Quantum Science and CAS Center for Excellence in Quantum Information and Quantum Physics,
		University of Science and Technology of China, Shanghai 201315, China}\par}

		\thispagestyle{titlepage}
	\end{center}
	\setcounter{equation}{0}
	\setcounter{figure}{0}
	\setcounter{table}{0}
	\setcounter{section}{0}
	\renewcommand{\theequation}{S\arabic{equation}}
	\renewcommand{\thefigure}{S\arabic{figure}}
	\renewcommand{\thetable}{S\arabic{table}}
	\renewcommand{\thesection}{S.\Roman{section}}

\renewcommand{\im}[1]{{\operatorname{im} \, #1}}
\renewcommand{\Ann}[1]{{\operatorname{Ann} \, #1}}
\let\figurename\SMfigurecaptionname
\let\tablename\SMtablecaptionname
\setcounter{thm}{0}
\setcounter{defn}{0}
\setcounter{conv}{0}
\setcounter{rem}{0}
\setcounter{lem}{0}
\setcounter{prop}{0}
\setcounter{cor}{0}
\setcounter{ex}{0}
\providecommand{\theHsection}{}
\renewcommand{\theHsection}{supp.\arabic{section}}
\providecommand{\theHsubsection}{}
\renewcommand{\theHsubsection}{\theHsection.\arabic{subsection}}
\providecommand{\theHsubsubsection}{}
\renewcommand{\theHsubsubsection}{\theHsubsection.\arabic{subsubsection}}
\providecommand{\theHequation}{}
\renewcommand{\theHequation}{supp.\arabic{equation}}
\providecommand{\theHfigure}{}
\renewcommand{\theHfigure}{supp.\arabic{figure}}
\providecommand{\theHtable}{}
\renewcommand{\theHtable}{supp.\arabic{table}}
\providecommand{\theHthm}{}
\renewcommand{\theHthm}{supp.\arabic{thm}}
\providecommand{\theHconv}{}
\renewcommand{\theHconv}{supp.\arabic{conv}}
\providecommand{\theHrem}{}
\renewcommand{\theHrem}{supp.\arabic{rem}}
\providecommand{\theHlem}{}
\renewcommand{\theHlem}{supp.\arabic{lem}}
\providecommand{\theHprop}{}
\renewcommand{\theHprop}{supp.\arabic{prop}}
\providecommand{\theHcor}{}
\renewcommand{\theHcor}{supp.\arabic{cor}}
\providecommand{\theHex}{}
\renewcommand{\theHex}{supp.\arabic{ex}}
\providecommand{\theHdefn}{}
\renewcommand{\theHdefn}{\theHsection.\arabic{defn}}

\section{Spacetime fault complex and gauge structure}
\label{sec:spacetime_complex_exactness}

This section develops the spacetime construction in the notation of the
main text.  The static differentials are denoted by $\partial_i$, the
spacetime fault differentials by $\Square_i$, and the completed
interaction-relation maps by $\widetilde{\Square}_i^\mu$.
Unless a finite quotient is specified, all configurations have finite
Laurent support on the infinite spatial or spacetime lattice.

\subsection{Static CSS complex and stabilizer relations}
\label{subsec:static_boundary_maps}

Let $R_{\rm s}=\mathbb F_2[x_1^{\pm1},\ldots,x_D^{\pm1}]$ be the
translation ring of a $D$-dimensional lattice.  A translation-invariant
CSS code with $q$ qubits, $a$ $X$-check types, and $b$ $Z$-check types per
unit cell is specified by stabilizer matrices
$A:R_{\rm s}^a\to R_{\rm s}^q$ and
$B:R_{\rm s}^b\to R_{\rm s}^q$~\cite{Haah13}.
We suppress overall Pauli phases and fix lattice bases, using the same
qubit ordering in the two Pauli sectors.  For a matrix $M$, the adjoint
is $M^\dagger=\bar M^{\mathsf T}$, with $\bar x_j=x_j^{-1}$; these bases
identify each free module with its coordinate dual.

The two static sector complexes are
\begin{equation}
 C_\bullet^X:\quad R_{\rm s}^a\xrightarrow{A}R_{\rm s}^q
                 \xrightarrow{B^\dagger}R_{\rm s}^b,
 \qquad
 C_\bullet^Z:\quad R_{\rm s}^b\xrightarrow{B}R_{\rm s}^q
                 \xrightarrow{A^\dagger}R_{\rm s}^a,
 \label{eq:SI_static_sectors}
\end{equation}
in degrees $1,0,-1$.  A sector label specifies the Pauli fault: an
$X$ fault violates $Z$ checks, and a $Z$ fault violates $X$ checks.
Thus $C_\bullet=C_\bullet^X\oplus C_\bullet^Z$ has check-label module
$C_1=R_{\rm s}^a\oplus R_{\rm s}^b$, Pauli-support module
$C_0=R_{\rm s}^q\oplus R_{\rm s}^q$, and syndrome module
$C_{-1}=R_{\rm s}^b\oplus R_{\rm s}^a$, with
\begin{equation}
 \partial_1=\begin{pmatrix}A&0\\0&B\end{pmatrix},
 \qquad
 \partial_0=\begin{pmatrix}B^\dagger&0\\0&A^\dagger\end{pmatrix}.
 \label{eq:SI_static_matrices}
\end{equation}
Check commutativity is $\partial_0\partial_1=0$.
The infinite-lattice topological condition is exactness at $C_0$:
\begin{equation}
 \ker\partial_0=\operatorname{im}\partial_1
 \quad\Longleftrightarrow\quad
 \ker B^\dagger=\operatorname{im}A,\qquad
 \ker A^\dagger=\operatorname{im}B.
 \label{eq:static_middle_exactness}
\end{equation}
It means that every finite-support Pauli operator with zero syndrome is
a product of stabilizers.  Both sector conditions are assumed below.

Choose a bounded finite-free extension by resolving the stabilizer
relations and their successive relations separately in each sector:
\begin{equation}
 C_\bullet:\quad\cdots\longrightarrow C_2
 \xrightarrow{\partial_2}C_1
 \xrightarrow{\partial_1}C_0
 \xrightarrow{\partial_0}C_{-1}\longrightarrow0,
 \qquad
 \ker\partial_i=\operatorname{im}\partial_{i+1}\quad(i\geq1).
 \label{eq:SI_static_complex}
\end{equation}
The higher differentials encode successive syzygies, as in the static
relation-layer framework in Ref.~\cite{Song26}.
We set $C_i=0$ for $i<-1$ and do not assume that $\partial_0$ is
surjective.  The sector differentials are written $\partial_i^\mu$.
For complementary sectors $\{\mu,\nu\}=\{X,Z\}$, the fixed bases give
$\partial_0^\nu=(\partial_1^\mu)^\dagger$ and
$\partial_1^\nu=(\partial_0^\mu)^\dagger$.
Here $C_0^\nu$ and $C_0^\mu$ have the same qubit-coordinate basis,
while $C_1^\nu$ and $C_{-1}^\mu$ have the same check-coordinate basis.
The higher relation modules are chosen independently in the two sectors.

\subsection{Temporal tensor product and spacetime exactness}
\label{subsec:spacetime_boundary_map}

Let $t$ translate by one measurement round and set
$R_{\rm st}=R_{\rm s}[t^{\pm1}]$.
Repeated syndrome extraction is encoded by the
one-generator temporal Koszul complex~\cite{Song26,Hillmann25}
\begin{equation}
 K_\bullet=
 \left[0\longrightarrow R_{\rm st}
 \xrightarrow{\,t-1\,}R_{\rm st}\longrightarrow0\right],
 \label{eq:SI_temporal_koszul}
\end{equation}
in degrees $1,0$.  The spacetime fault complex is its total tensor
product with the static complex:
\begin{equation}
 F_\bullet=C_\bullet\otimes_{R_{\rm s}}K_\bullet,
 \qquad
 F_i=(C_i\oplus C_{i-1})\otimes_{R_{\rm s}}R_{\rm st}.
 \label{eq:SI_temporal_tensor_product}
\end{equation}
The two summands have temporal degrees $0,1$, respectively.  Over
$\mathbb F_2$, the tensor-product signs disappear, giving
\begin{equation}
 \Square_i=
 \begin{pmatrix}\partial_i&(t-1)I\\0&\partial_{i-1}\end{pmatrix}
 \quad(i\geq1),
 \qquad
 \Square_0=\begin{pmatrix}\partial_0&(t-1)I\end{pmatrix}.
 \label{eq:general_spacetime_resolution}
\end{equation}
All degrees of the tensor product are retained, including the last
temporal summand above the highest nonzero static module.
Scalar extension to $R_{\rm st}$ is implicit from now on: for example,
$A$ denotes $A\otimes_{R_{\rm s}}\mathrm{id}_{R_{\rm st}}$
as a map $R_{\rm st}^a\to R_{\rm st}^q$, with unchanged matrix entries.
The same convention applies to the other static maps, and the
adjoint also sends $t\mapsto t^{-1}$.

The relevant part of the fault complex is
\begin{equation}
 F_2\xrightarrow{\Square_2}
 F_1\xrightarrow{\Square_1}
 F_0\xrightarrow{\Square_0}F_{-1}\longrightarrow0.
 \label{eq:spacetime_fault_complex}
\end{equation}
Each module is defined by Eq.~\eqref{eq:SI_temporal_tensor_product};
in particular, $F_{-1}=C_{-1}\otimes_{R_{\rm s}}R_{\rm st}$ because
$C_{-2}=0$.
For a fault history $\xi=(e,m)\in F_0$, the detector output is
$\Square_0\xi=\partial_0e+(t-1)m$, where $e$ records data faults and
$m$ records measurement faults.  The map $\Square_1$ generates local
fault-history deformations, and $\Square_2$ generates relations among
those deformations.  The chain identities
$\Square_0\Square_1=0$ and $\Square_1\Square_2=0$ follow from
$\partial_{i-1}\partial_i=0$.  In particular, all histories in
$[\xi]=\xi+\operatorname{im}\Square_1$ have the same detector output.

The CSS splitting is preserved:
$F_\bullet=F_\bullet^X\oplus F_\bullet^Z$, with
$F_\bullet^\mu=C_\bullet^\mu\otimes_{R_{\rm s}}K_\bullet$ and
sector maps $\Square_i^\mu$.
For example,
\begin{equation}
 \Square_1^X=\begin{pmatrix}A&(t-1)I_q\\0&B^\dagger\end{pmatrix},
 \qquad
 \Square_1^Z=\begin{pmatrix}B&(t-1)I_q\\0&A^\dagger\end{pmatrix}.
 \label{eq:SI_sector_fault_maps}
\end{equation}
This algebraic decomposition does not assume statistically independent
noise in the two sectors.

\begin{prop}[Temporal homology preservation]
\label{prop:SI_temporal_homology}
For the Laurent-polynomial complexes above,
\begin{equation}
 H_i(F_\bullet)\cong H_i(C_\bullet)
 \label{eq:SI_homology_preservation}
\end{equation}
as $R_{\rm s}$-modules, with temporal translation acting as the identity
on the right-hand side.  The same statement holds sectorwise.
\end{prop}
\begin{proof}
Multiplication by $t-1$ is injective, so $K_\bullet$ is a free resolution
of $R_{\rm st}/(t-1)\cong R_{\rm s}$; this is the one-generator case
of the Koszul-resolution property~\cite[Corollary~4.5.5]{Weibel94}.
Tensoring its augmentation with the bounded free complex $C_\bullet$
preserves the quasi-isomorphism.  The resulting chain map is
$(u,v)\mapsto u|_{t=1}$ and induces the stated homology isomorphism.
\end{proof}
Evaluation at $t=1$ sums temporal coefficients; it does not select a
physical measurement slice.  Static exactness therefore gives
\begin{equation}
 \ker\Square_0^\mu=\operatorname{im}\Square_1^\mu,
 \qquad
 \ker\Square_i^\mu=\operatorname{im}\Square_{i+1}^\mu\quad(i\geq1).
 \label{eq:gauge_side_completeness}
\end{equation}
Every finite-support detector-free history is consequently a finite
combination of translated local deformations, and the higher maps
generate all their finite-support relations.  This statement does not
bound the spatial extent of a filling.  Periodic time can introduce
wrapping classes, and finite open-time protocols require their own
boundary maps.

\begin{cor}[Periodic-family fault distance]
\label{cor:periodic_fault_distance}
For periodic spacetime quotients with diverging injectivity radius,
fixed finite-range maps, and an exact infinite-lattice sector complex,
\begin{equation}
 d_{{\rm f},\mu}(L)\coloneqq
 \min\left\{|\xi|:\xi\in\ker\Square_{0,L}^\mu
          \setminus\operatorname{im}\Square_{1,L}^\mu\right\}
 \longrightarrow\infty.
 \label{eq:periodic_fault_distance}
\end{equation}
Here $|\xi|$ counts sector fault locations, and an empty minimum is
$\infty$.
\end{cor}
\begin{proof}
Otherwise there are nontrivial detector-free histories of bounded
weight.  Connect two fault locations when their detector supports
overlap.  Each connected component has bounded diameter and is itself
detector free; at least one component is nontrivial.  Up to translation
there are finitely many such patterns, so a subsequence lifts to a fixed
finite-support history on the infinite lattice.  Exactness supplies a
finite-support filling, which fits in an injectivity ball for all
sufficiently large quotients.  Projecting that filling makes the chosen
component trivial, a contradiction.
\end{proof}
The conclusion is distance divergence.  Its scaling rate and the
numerical decoding threshold require further information about the code
family and the noise model.

\subsection{Dual complex and gauge theory}
\label{subsec:SI_gauge_model}

The adjoint of the fault complex is the cochain complex
\begin{equation}
 (F_\bullet^\mu)^\dagger:\quad
 F_{-1}^\mu\xrightarrow{(\Square_0^\mu)^\dagger}
 F_0^\mu\xrightarrow{(\Square_1^\mu)^\dagger}
 F_1^\mu\xrightarrow{(\Square_2^\mu)^\dagger}
 F_2^\mu\longrightarrow\cdots .
 \label{eq:SI_dual_fault_complex}
\end{equation}
The maps $(\Square_1^\mu)^\dagger$ and $(\Square_2^\mu)^\dagger$
encode interaction supports and gauge-charge patterns, respectively.
Finite freeness gives the tensor-product duality
\begin{equation}
 (F_\bullet^\mu)^\dagger
 =(C_\bullet^\mu\otimes_{R_{\rm s}}K_\bullet)^\dagger
 \cong(C_\bullet^\mu)^\dagger\otimes_{R_{\rm s}}K_\bullet^\dagger,
 \label{eq:SI_tensor_adjoint_duality}
\end{equation}
with total tensor products and adjoints over $R_{\rm s}$ and
$R_{\rm st}$ on the two factors on the right.
The temporal adjoint has differential $t^{-1}-1$ and cohomology
$R_{\rm s}$ only in degree $1$, so
Eq.~\eqref{eq:SI_tensor_adjoint_duality} gives
\begin{equation}
 H^j\bigl((F_\bullet^\mu)^\dagger\bigr)
 \cong H^{j-1}\bigl((C_\bullet^\mu)^\dagger\bigr),
 \qquad j\in\mathbb Z,
 \label{eq:SI_dual_homology_preservation}
\end{equation}
on the infinite lattice, as $R_{\rm s}$-modules with $t$ acting as the
identity on the right.

On a translation-invariant system, introduce auxiliary-spin coordinates $s\in F_1^\mu$, with
$\sigma_j=(-1)^{s_j}$.  For a spin support $\Sigma\in F_1^\mu$, set
$W_\Sigma(s)=(-1)^{\langle\Sigma,s\rangle}$,
where the pairing is the binary dot product.
With unit coupling magnitude, the boundary-map construction gives
\begin{equation}
 H_{\xi_\mu}(s)
 =-\sum_\ell J_\ell
       W_{(\Square_1^\mu)^\dagger\ell}(s),
 \qquad J_\ell=(-1)^{\xi_\mu(\ell)}.
 \label{eq:SI_decoding_hamiltonian}
\end{equation}
Here $\ell$ is a fault-location basis vector in $F_0^\mu$, and
$\xi_\mu$ fixes the quenched signs.  Since
$\langle(\Square_1^\mu)^\dagger\ell,s\rangle
=(\Square_1^\mu s)(\ell)$, the spins parametrize the fault histories
$\xi_\mu+\Square_1^\mu s$.
For independent binary faults of common rate $p$, the partition
function on the Nishimori line $e^{-2\beta}=p/(1-p)$ is proportional
to the probability of $[\xi_\mu]$.  The normalization, including the
uniform multiplicity $|\ker\Square_1^\mu|$, is derived in
Sec.~\ref{subsec:statistical_mechanical_mapping}.

The Hamiltonian in Eq.~\eqref{eq:SI_decoding_hamiltonian} defines a
gauge theory with gauge transformations of the form
$s\mapsto s+\Square_2^\mu\chi$.
These preserve every interaction because
$\Square_1^\mu\Square_2^\mu=0$.
By Eq.~\eqref{eq:gauge_side_completeness}, these shifts generate all
finite-support interaction-preserving spin flips on the infinite
lattice.  A monomial is invariant under these gauge transformations
precisely when
\begin{equation}
 (\Square_2^\mu)^\dagger\Sigma=0.
 \label{eq:SI_wilson_neutrality}
\end{equation}
The vector $(\Square_2^\mu)^\dagger\Sigma$ is its gauge-charge pattern.
Neutral supports modulo supports generated by interactions form
$H^1\bigl((F_\bullet^\mu)^\dagger\bigr)$.
On the infinite lattice, the $j=1$ case of
Eq.~\eqref{eq:SI_dual_homology_preservation} and the CSS pairing
identities identify this quotient with
$H_0(C_\bullet^\nu)$ for the complementary sector $\nu$.
Static exactness therefore gives
\begin{equation}
 H^1\bigl((F_\bullet^\mu)^\dagger\bigr)=0,
 \qquad
 \ker(\Square_2^\mu)^\dagger
 =\operatorname{im}(\Square_1^\mu)^\dagger.
 \label{eq:dagger_side_exactness}
\end{equation}
Thus $(\Square_1^\mu)^\dagger$ generates all finite-support neutral
supports on the infinite lattice, so their monomials are products of
interactions.  Finite periodic quotients may retain additional global
neutral classes.

See Section~\ref{sec:code_induced_gauge_theories} for concrete examples of order-parameter construction.

\subsection{Resolution of interaction relations}
\label{subsec:chain_resolutions}

Summing over auxiliary spins selects products of interactions with
zero total support.  Such an interaction relation is specified by
$g\in F_0^\mu$ satisfying $(\Square_1^\mu)^\dagger g=0$.
The detector adjoint $(\Square_0^\mu)^\dagger$ generates such graphs,
but can miss classes.  On the infinite lattice,
Eq.~\eqref{eq:SI_dual_homology_preservation} and CSS pairing identify
the obstruction:
\begin{equation}
 \frac{\ker(\Square_1^\mu)^\dagger}
      {\operatorname{im}(\Square_0^\mu)^\dagger}
 \cong H^{-1}\bigl((C_\bullet^\mu)^\dagger\bigr)
 =\ker_{R_{\rm s}}\partial_1^\nu.
 \label{eq:SI_adjoint_obstruction}
\end{equation}
The missing classes therefore come from static stabilizer relations
in the complementary sector.

With the fixed coordinate bases, we identify
$C_0^\nu=(C_0^\mu)^\dagger$ and
$C_{\pm1}^\nu=(C_{\mp1}^\mu)^\dagger$,
with differentials related by the CSS pairing identities.
The higher static relation modules are chosen independently.

To resolve the interaction relations, tensor $C_\bullet^\nu$ with
a temporal factor whose differential is $t^{-1}-1$, in homological
degrees $1,0$.  Denote the resulting tensor-product complex by
$\widetilde F_\bullet^\nu$.  Listing the temporal-degree-$1$ summand
first gives the modules
$\widetilde F_i^\nu=(C_{i-1}^\nu\oplus C_i^\nu)
\otimes_{R_{\rm s}}R_{\rm st}$
and the differentials
\begin{equation}
 \widetilde{\Square}_i^\nu
 \coloneqq\begin{pmatrix}
   \partial_{i-1}^\nu&0\\(t^{-1}-1)I&\partial_i^\nu
  \end{pmatrix},\qquad i\geq1.
 \label{eq:higher_cross_sector_correspondence}
\end{equation}
The CSS pairing identities give
$\widetilde F_i^\nu\cong(F_{1-i}^\mu)^\dagger$ for $i=0,1$.
Thus chain degrees $1,0$ correspond to adjoint cochain degrees $0,1$.
Under the fixed coordinate-dual identifications, the central map is
$\widetilde{\Square}_1^\nu=(\Square_1^\mu)^\dagger$.
At $i=2$, however,
$\widetilde F_2^\nu\cong(F_{-1}^\mu)^\dagger
\oplus(C_2^\nu\otimes_{R_{\rm s}}R_{\rm st})$;
the additional summand supplies the static relation generators:
\begin{equation}
 \widetilde{\Square}_2^\nu
 =\left((\Square_0^\mu)^\dagger\quad
         \begin{pmatrix}0\\\partial_2^\nu\end{pmatrix}\right).
 \label{eq:SI_completed_graph_map}
\end{equation}
Its source is $\widetilde F_2^\nu$, and its target is
$\widetilde F_1^\nu\cong F_0^\mu$.
The completed interaction complex is
\begin{equation}
 \cdots\xrightarrow{\widetilde{\Square}_3^\nu}
 \widetilde F_2^\nu\xrightarrow{\widetilde{\Square}_2^\nu}
 \bigl(\widetilde F_1^\nu\cong F_0^\mu\bigr)
 \xrightarrow{\widetilde{\Square}_1^\nu=(\Square_1^\mu)^\dagger}
 \bigl(\widetilde F_0^\nu\cong F_1^\mu\bigr)
 \xrightarrow{(\Square_2^\mu)^\dagger}
 F_2^\mu\xrightarrow{(\Square_3^\mu)^\dagger}\cdots .
 \label{eq:SI_paired_interaction_segment}
\end{equation}
Here, the complementary-sector resolution is spliced into the ordinary adjoint at $F_0^\mu$, supplying the incoming relation maps while retaining $(\Square_1^\mu)^\dagger$ and the subsequent adjoint maps.

\begin{prop}[Complete graph generation and higher relations]
\label{prop:SI_graph_completeness}
For complementary CSS sectors on the infinite spacetime lattice,
\begin{equation}
 \operatorname{im}\widetilde{\Square}_{i+1}^\nu
 =\ker\widetilde{\Square}_i^\nu,\qquad i\geq1.
 \label{eq:graph_completeness}
\end{equation}
\end{prop}
\begin{proof}
Since $t^{-1}-1=t^{-1}(t-1)$ over $\mathbb F_2$, the temporal factor
is again a free resolution of $R_{\rm s}$.
The argument of Proposition~\ref{prop:SI_temporal_homology} therefore
gives $H_i(\widetilde F_\bullet^\nu)\cong H_i(C_\bullet^\nu)=0$
for $i\geq1$.
\end{proof}
The case $i=1$ gives
$\operatorname{im}\widetilde{\Square}_2^\nu
=\ker(\Square_1^\mu)^\dagger$.
Thus the complementary tail is a free resolution of the
interaction-relation module, without requiring code self-duality.
Higher cohomology of the ordinary adjoint may still be nonzero.
On a finite periodic quotient, spin summation still selects the full
kernel, but global graphs need not be images of the reduced local
generator map.

\section{Boundary-Generated Threshold Computation}
\label{sec:boundary_generated_threshold_computation}

In this section, we review the standard statistical-mechanical mapping and the connection between code correctability and phase transitions~\cite{Dennis02,Song22}.
We then present our algebraic machinery based on graph expansions, which operates directly on the boundary maps of the interaction complex, without requiring an explicit construction of the spin Hamiltonian.

\subsection{Statistical-Mechanical Mapping}
\label{subsec:statistical_mechanical_mapping}

Recovering encoded quantum information requires identifying the correct fault-equivalence class.  
For independent data and measurement faults of rate $p$,
a fault history $\xi_\mu$ has probability
$\operatorname{Pr}(\xi_\mu;p)
=p^{|\xi_\mu|}(1-p)^{N_{\rm f}-|\xi_\mu|}$, where $N_{\rm f}$ counts
the fault locations.  
Optimal decoding compares the total probabilities
of classes $[\xi_\mu]=\xi_\mu+\operatorname{im}\Square_1^\mu$,
\begin{equation}
 \operatorname{Pr}([\xi_\mu];p)
 =\sum_{\xi'_\mu\in[\xi_\mu]}\operatorname{Pr}(\xi'_\mu;p).
 \label{eq:fault_class_probability}
\end{equation}

Let $\rho$ be an arbitrary encoded state and $E_{\xi_\mu}$ the
accumulated data-error Pauli operator. 
Fault histories in one class have the same detector record and act identically on the code space. 
For a record $\mathbf d=\Square_0^\mu\xi_\mu$, choose a reference history
$\xi_{\mathbf d}$.  The logical alternatives are
$[\xi_{\mathbf d}+\lambda]$, where $\lambda$ runs over detector-free
histories representing inequivalent logical operators.
The noisy state can
therefore be grouped by fault class, or equivalently by record and
logical label:
\begin{align}
 \rho_1
 &=\sum_{\xi_\mu}\operatorname{Pr}(\xi_\mu;p)
       E_{\xi_\mu}\rho E_{\xi_\mu}^\dagger\nonumber\\
 &=\sum_{\mathbf d,\lambda}
       \operatorname{Pr}([\xi_{\mathbf d}+\lambda];p)
       E_{\xi_{\mathbf d}+\lambda}\rho
       E_{\xi_{\mathbf d}+\lambda}^\dagger.
 \label{eq:fault_class_noisy_state}
\end{align}

Optimal decoding chooses a class with the largest total
probability for each record.  Denote its representative by
$\xi_{\mathbf d}^*$ and take this class as the origin
$\lambda=0$.  Applying $E_{\xi_{\mathbf d}^*}^\dagger$
restores the encoded state if the selected class is correct; otherwise
the residual action is the logical operator $E_\lambda$.
Averaging over records, the recovered state is
\begin{align}
 \rho_2
 &=P_{\rm succ}\rho
   +\sum_{\mathbf d}\sum_{\lambda\ne0}
       \operatorname{Pr}([\xi_{\mathbf d}^*+\lambda];p)
       E_\lambda\rho E_\lambda^\dagger,\nonumber\\
 P_{\rm succ}
 &=\sum_{\mathbf d}\operatorname{Pr}([\xi_{\mathbf d}^*];p).
 \label{eq:maximum_likelihood_recovery}
\end{align}
The encoded information is recoverable if $P_{\rm succ}\to1$ as the code size grows, namely, the total probability
of all other logical sectors vanish.

The histories in $[\xi_\mu]$ are parametrized by
$\xi_\mu+\Square_1^\mu s$, with $s\in F_1^\mu$.
Each history has $|\ker\Square_1^\mu|$ preimages, so summing over $s$
counts every history with the same multiplicity.
Introduce $\beta$ through the Nishimori condition
\begin{equation}
 e^{-2\beta}=\frac{p}{1-p}.
 \label{eq:Nishimori}
\end{equation}
Using $p^x(1-p)^{1-x}=[p(1-p)]^{1/2}e^{\beta(-1)^x}$, 
we can rewrite the total class probability as
\begin{equation}
 \operatorname{Pr}([\xi_\mu];p)
 \propto\mathcal Z_{\xi_\mu}(\beta)
 =\sum_{s\in F_1^\mu}
 \exp\!\left[\beta\sum_\ell
 (-1)^{\xi_\mu(\ell)+(\Square_1^\mu s)(\ell)}\right].
 \label{eq:general_partition_start}
\end{equation}
Here $\mathcal Z_{\xi_\mu}(\beta)$ is the partition function.
The proportionality factor $[p(1-p)]^{N_{\rm f}/2}/|\ker\Square_1^\mu|$
is the same for every class and cancels in their probability ratios.
Identifying the exponent in Eq.~\eqref{eq:general_partition_start} as $-\beta H_{\xi_\mu}(s)$ recovers the auxiliary-spin Hamiltonian in Eq.~\eqref{eq:SI_decoding_hamiltonian}.

Thus, the ratio between two classes compatible with the same record is given by
\begin{equation}
 \frac{\operatorname{Pr}([\xi_{\mathbf d}^*+\lambda];p)}
      {\operatorname{Pr}([\xi_{\mathbf d}^*];p)}
 =\frac{\mathcal Z_{\xi_{\mathbf d}^*+\lambda}(\beta)}
       {\mathcal Z_{\xi_{\mathbf d}^*}(\beta)}
 =e^{-\beta\Delta\mathcal F_{\mathbf d,\lambda}}.
 \label{eq:logical_class_free_energy_cost}
\end{equation}
Here $\Delta\mathcal F_{\mathbf d,\lambda}$ is the free-energy cost of inserting a logical defect, without assuming a particular defect geometry.
In the ordered phase, the relative probabilities of logical defects are
exponentially suppressed.  The code remains correctable when the total
probability of nontrivial logical sectors vanishes as the code size grows,
giving $P_{\rm succ}\to1$.

\subsection{Boundary-generated graph expansion}
\label{subsec:boundary_expansion}

To locate the phase transition, we compute the quenched free energy
through a graph expansion generated directly from the boundary maps.
As quenched disorder leaves graph generation unchanged, we first derive
the expansion at fixed disorder, then extract irreducible contributions
and carry out the disorder average analytically.

For the partition function in
Eq.~\eqref{eq:general_partition_start}, set $v=\tanh\beta$ and write
$S_\ell=\prod_{i\in\supp((\Square_1^\mu)^\dagger\ell)}\sigma_i$.
Expanding the product over the interaction set $\mathcal I_\mu$ using
$e^{\beta J_\ell S_\ell}=\cosh\beta(1+vJ_\ell S_\ell)$
labels each term by a selected interaction set
$g\subseteq\mathcal I_\mu$.
The spin sum retains a term
precisely when $(\Square_1^\mu)^\dagger g=0$, since every auxiliary spin
must occur an even number of times. Hence
\begin{equation}
    \mathcal Z_{\xi_\mu}(\beta)
    =
    Z_0(\beta)
    \sum_{\substack{g\subseteq\mathcal I_\mu\\
                    (\Square_1^\mu)^\dagger g=0}}
    v^{|g|}\prod_{\ell\in g}J_\ell .
    \label{eq:general_partition_kernel_form}
\end{equation}
Here $Z_0(\beta)=2^{N_\sigma}(\cosh\beta)^{N_{\rm f}}$,
$N_\sigma$ is the number of auxiliary spins,
$N_{\rm f}=|\mathcal I_\mu|$, and $|g|$ counts selected interactions.
Let $N_{\rm fr}$ denote the number of unit cells.

We construct the filling expansion on the infinite lattice for a
fixed CSS sector $\mu$ and keep its complete generator set and
maps fixed. A generator support $\gamma$ is a finite set of generator
placements, each labeled by position and type; $|\gamma|$ counts
its generators.
We call $\gamma$ a filling of its boundary $g=\widetilde{\Square}_2^\nu\gamma$.

Equation~\eqref{eq:graph_completeness} ensures that every admissible
graph of finite support on this lattice has a finite filling.
The map $\widetilde{\Square}_2^\nu$ may have a nontrivial kernel, so a boundary can admit
distinct fillings. These are retained as separate configurations.

To count each boundary once when summing over all its fillings,
we distribute its graph weight uniformly over its global minimum
fillings. For any boundary
$g\in\operatorname{im}\widetilde{\Square}_2^\nu$, define
\begin{equation}
    m_{\min}(g)=\min_{\widetilde{\Square}_2^\nu\gamma=g}|\gamma|,
    \qquad
    \mathcal M(g)
    =\{\gamma:\widetilde{\Square}_2^\nu\gamma=g,\ |\gamma|=m_{\min}(g)\}.
    \label{eq:global_minimum_fillings}
\end{equation}
The minimum is taken over all finite fillings in the complete
infinite system.
In particular, $m_{\min}(0)=0$ and $\mathcal M(0)=\{\varnothing\}$.

For each boundary $g\in\operatorname{im}\widetilde{\Square}_2^\nu$,
the set of global minimum fillings $\mathcal M(g)$ is
finite.
We define the filling probabilities by
\begin{equation}
    \rho(\gamma)=
    \begin{cases}
        |\mathcal M(\widetilde{\Square}_2^\nu\gamma)|^{-1},
        & \gamma\in\mathcal M(\widetilde{\Square}_2^\nu\gamma),\\
        0, & \text{otherwise}.
    \end{cases}
    \label{eq:global_filling_probability}
\end{equation}
The probabilities are normalized separately for each boundary
$g\in\operatorname{im}\widetilde{\Square}_2^\nu$,
$\sum_{\widetilde{\Square}_2^\nu\gamma=g}\rho(\gamma)=1$.

These probabilities are independent of $p$, $\beta$, and the disorder
realization. For $\gamma'\subseteq\gamma$, both minimality and the
number of minimum fillings of $\widetilde{\Square}_2^\nu\gamma'$ are determined globally,
including competing fillings outside $\gamma$. The same
$\rho(\gamma')$ is used on every support without local renormalization.

At fixed disorder, write $\psi(g)=v^{|g|}\prod_{\ell\in g}J_\ell$
for the graph factor. For a finite generator support $\gamma$,
define the local weighted polynomial
\begin{equation}
    \mathcal P(\gamma)
    =\sum_{\gamma'\subseteq\gamma}
    \rho(\gamma')\psi(\widetilde{\Square}_2^\nu\gamma').
    \label{eq:minimum_filling_local_polynomial}
\end{equation}
Each boundary $g$ contributes $\psi(g)$ times the fraction of all
its global minimum fillings contained in $\gamma$.

The polynomial $\mathcal P(\gamma)$ has constant term $1$, arising
from the empty filling. We define the local logarithm
$\mathcal L(\gamma)=\log\mathcal P(\gamma)$ as a formal power
series in $v$.

We decompose the local logarithm into contributions associated with
generator supports. Assigning an irreducible weight
$\mathcal W(\gamma')$ to each subset $\gamma'\subseteq\gamma$,
we require
$\mathcal L(\gamma)=\sum_{\gamma'\subseteq\gamma}\mathcal W(\gamma')$
for every finite $\gamma$.
The weight on $\gamma$ is then the contribution remaining after
those of its proper subsupports have been subtracted:
\begin{equation}
    \mathcal W(\gamma)
    =\mathcal L(\gamma)
     -\sum_{\gamma'\subsetneq\gamma}\mathcal W(\gamma').
    \label{eq:minimum_filling_mobius_weight}
\end{equation}
The sum runs over all proper subsets of $\gamma$, including
disconnected ones.

To evaluate $\mathcal W(\gamma)$ explicitly, we group products in
the logarithmic expansion by unordered multisets of fillings
$\mathcal A=\{\gamma_1,\ldots,\gamma_r\}$, $r\ge1$,
with $\gamma_i\subseteq\gamma$ and $\widetilde{\Square}_2^\nu\gamma_i\ne0$.
Repetitions are allowed. Write $r_{\gamma'}$ for the multiplicity of
each distinct filling; products over $\gamma'$ below run over these
distinct fillings. Combining the logarithmic coefficient with the
number of orderings gives
$c_{\mathcal A}=(-1)^{r+1}(r-1)!/\prod_{\gamma'}r_{\gamma'}!$.

Define the occupied support by the ordinary union
$\operatorname{Occ}(\mathcal A)=\bigcup_{i=1}^r\gamma_i$.
The term labeled by $\mathcal A$ occurs in the local logarithm of
every support containing this occupied support, with the same
coefficient and filling probabilities. The recursion in
Eq.~\eqref{eq:minimum_filling_mobius_weight} assigns this term to
$\mathcal W(\operatorname{Occ}(\mathcal A))$ and cancels its
contribution to the weights of all proper supersets of this
occupied support. Hence only multisets occupying all of $\gamma$
contribute:
\begin{equation}
    \mathcal W(\gamma)
    =\sum_{\substack{\mathcal A\\
                    \operatorname{Occ}(\mathcal A)=\gamma}}
      c_{\mathcal A}
      \prod_{\gamma'}
      \left[\rho(\gamma')\psi(\widetilde{\Square}_2^\nu\gamma')\right]^{r_{\gamma'}}.
    \label{eq:minimum_filling_occ_identity}
\end{equation}
Nonempty supports with $\widetilde{\Square}_2^\nu\gamma=0$ are included: subfillings
with nonzero boundaries may still cover $\gamma$.

For the disorder average, we sample physical fault histories from
the independent noise model of
Sec.~\ref{subsec:statistical_mechanical_mapping}, without conditioning
on a fixed detection record. The signs at distinct interaction
locations are independent, with $\Pr(J_\ell=-1)=p$ and
$\Pr(J_\ell=+1)=1-p$; set $q_{\mathrm{dis}}=1-2p$.
The quenched average is taken after the local logarithm and the
irreducible weights have been constructed at fixed disorder.
The coefficients $c_{\mathcal A}$ and filling probabilities $\rho$
are independent of the disorder realization, so the average acts
only on the product of graph factors in
Eq.~\eqref{eq:minimum_filling_occ_identity}.

For a multiset $\mathcal A$, boundaries of different fillings may
overlap, and repeated fillings reuse the same interaction signs.
Write the complete product as
$\prod_{\gamma'}\psi(\widetilde{\Square}_2^\nu\gamma')^{r_{\gamma'}}
=\prod_\ell(J_\ell v)^{n_\ell(\mathcal A)}$,
where $n_\ell(\mathcal A)=\sum_{\gamma'}r_{\gamma'}(\widetilde{\Square}_2^\nu\gamma')_\ell$
counts the occurrences of location $\ell$ over all fillings, with
the sum taken over the integers.
These occurrences must be combined before averaging because
they involve the same random sign.
For example, two occurrences give $[J_\ell^2]_{\mathrm{dis}}=1$,
while their temperature factor remains $v^2$.

For an integer multiplicity $n_\ell$, the single-sign moment is
$[J_\ell^{n_\ell}]_{\mathrm{dis}}=(1-p)+p(-1)^{n_\ell}$:
it equals $1$ for even $n_\ell$ and $q_{\mathrm{dis}}$ for odd $n_\ell$.
Let $k(\mathcal A)$ count the locations with odd multiplicity, and
let $m(\mathcal A)=\sum_\ell n_\ell(\mathcal A)
=\sum_{i=1}^r|\widetilde{\Square}_2^\nu\gamma_i|$ be the total interaction multiplicity.
Independence between distinct locations then gives
\begin{equation}
    \left[\prod_\ell(J_\ell v)^{n_\ell(\mathcal A)}\right]_{\mathrm{dis}}
    =v^{m(\mathcal A)}\prod_\ell
      \left[J_\ell^{n_\ell(\mathcal A)}\right]_{\mathrm{dis}}
    =q_{\mathrm{dis}}^{k(\mathcal A)}v^{m(\mathcal A)}.
    \label{eq:minimum_filling_disorder_moment}
\end{equation}
Substituting this result into
Eq.~\eqref{eq:minimum_filling_occ_identity} gives the quenched
irreducible weight
$\mathcal W_{\mathrm{av}}(\gamma;q_{\mathrm{dis}},v)
=[\mathcal W(\gamma)]_{\mathrm{dis}}$:
\begin{equation}
    \mathcal W_{\mathrm{av}}(\gamma;q_{\mathrm{dis}},v)
    =\sum_{\substack{\mathcal A\\
                    \operatorname{Occ}(\mathcal A)=\gamma}}
      c_{\mathcal A}
      \left[\prod_{\gamma'}\rho(\gamma')^{r_{\gamma'}}\right]
      q_{\mathrm{dis}}^{k(\mathcal A)}v^{m(\mathcal A)}.
    \label{eq:minimum_filling_disorder_weight}
\end{equation}

For the periodic finite-range maps above, with finitely many types per
unit cell and the exactness in Eq.~\eqref{eq:graph_completeness}, the
global minimum fillings of any boundary $g$ with $|g|\le N$ have bounded
size and lie within a bounded neighborhood of $g$. Both bounds depend
only on $N$ and the fixed lattice maps. Each filling in a nonzero
multiset term contributes at least one power of $v$, so a term through
order $v^N$ contains at most $N$ global minimum fillings. Their union
has bounded size, so supports containing sufficiently many generators
make no contribution through order $v^N$. For each fixed finite support,
computing its weight through this order requires only finitely many
multiset terms.

A bound on support size still allows its parts to be arbitrarily far
apart, so their separation must also be controlled. If a support splits
into two nonempty parts separated by more than a distance depending on
$N$, all global minimum fillings of a relevant boundary with $|g|\le N$
are obtained by choosing a minimum filling independently for each
boundary part and taking their union. The numbers of choices multiply,
so the uniform filling probabilities also multiply. Through order $v^N$,
the local polynomial on the full support or any of its subsets is
therefore the product of the corresponding polynomials on the two sides.
Taking the logarithm turns these products into sums, and the recursion
in Eq.~\eqref{eq:minimum_filling_mobius_weight} cancels the weight of
the full support through this order. Together, these size and separation
restrictions leave only finitely many contributing supports up to
translation.

Let $\mathfrak G_{\mathrm{tr}}$ contain one representative of each
translation class of nonempty finite generator supports and define
the coefficients per unit cell by
\begin{equation}
    A_{m,k}
    =\sum_{\gamma\in\mathfrak G_{\mathrm{tr}}}
      [q_{\mathrm{dis}}^k v^m]\,\mathcal W_{\mathrm{av}}(\gamma;q_{\mathrm{dis}},v).
    \label{eq:minimum_filling_series_coefficients}
\end{equation}

Since the filling probabilities sum to one for each boundary, summing
over all global minimum fillings of an admissible finite graph counts
its graph factor exactly once. The local logarithms and irreducible
weights are constructed at fixed disorder before averaging. Together
with the finite spatial extent of the contributions at each fixed order,
this identifies $A_{m,k}$ as the coefficients of $[\log(\mathcal
Z_{\xi_\mu}/Z_0)]_{\mathrm{dis}}$ per unit cell in the thermodynamic
limit. Restoring the prefactor $Z_0$ gives the following expansion of
the quenched free-energy density, understood coefficientwise about
$v=0$:
\begin{equation}
    \begin{aligned}
        \beta[f_{\xi_\mu}]_{\mathrm{dis}}
        &\coloneqq
        -\lim_{N_{\rm fr}\to\infty}
        \frac{1}{N_{\rm fr}}
        \left[\log\mathcal Z_{\xi_\mu}(\beta)\right]_{\mathrm{dis}}\\
        &=\beta f_0
          -\sum_{m=1}^{\infty}\sum_{k=0}^m A_{m,k}q_{\mathrm{dis}}^k v^m.
    \end{aligned}
    \label{eq:minimum_filling_free_energy}
\end{equation}
Here $\beta f_0=-\lim_{N_{\rm fr}\to\infty}
N_{\rm fr}^{-1}\log Z_0(\beta)$ is the contribution of the
disorder-independent prefactor $Z_0$ and is analytic at finite
positive temperature.

\section{Code Examples and Stabilizer Maps}
	\label{sec:quantum_codes_studied}

	We specify the six code examples using the conventions of
	Sec.~\ref{sec:spacetime_complex_exactness}, writing $R=R_{\rm s}$.
	For a Laurent polynomial $h$, $h^\ast=h(x_1^{-1},\ldots,x_D^{-1})$
	denotes spatial inversion.
	The columns of the complete stabilizer map list the $Z$-checks followed
	by the $X$-checks.  Its first $q$ rows record $Z$-type Pauli support
	and its last $q$ rows record $X$-type support:
	\begin{equation}
		\mathsf S_{\rm code}
		=
		\begin{pmatrix}
			B&0\\
			0&A
		\end{pmatrix}.
		\label{eq:code_stabilizer_map_convention}
	\end{equation}
	The stabilizer group is generated by the Pauli operators encoded by
	these columns and all their lattice translates.

	\subsection{Fracton codes}
	\label{subsec:fracton_codes}

	We consider three prominent fracton codes: Haah's code, the checkerboard code, and the X-cube
	code.

	\paragraph{Haah's cubic code.}
	Haah's code has two qubits at each vertex of a simple-cubic lattice,
	with one $X$-check and one $Z$-check per primitive unit
	cell~\cite{Haah11,Vijay16}.  We denote unit translations by $x,y,z$
	and order the qubit species as $(1,2)$.  With $f=1+x+y+z$ and
	$g=1+xy+yz+zx$, the complete stabilizer map is
	\begin{equation}
		\mathsf S_{\rm Haah}
		=
		\begin{pmatrix}
			g^\ast&0\\
			f^\ast&0\\
			0&f\\
			0&g
		\end{pmatrix}
		=
		\begin{pmatrix}
			1+\bar x\bar y+\bar y\bar z+\bar z\bar x&0\\
			1+\bar x+\bar y+\bar z&0\\
			0&1+x+y+z\\
			0&1+xy+yz+zx
		\end{pmatrix}.
		\label{eq:haah_code_stabilizer_map}
	\end{equation}
	Both $X$- and $Z$-type checks have weight eight.

	Haah's code has point excitations created at the corners of fractal
	Pauli operators.  No finite-width string operator can create an
	isolated pair of widely separated excitations, a hallmark of Type-II
	fracton order.

	\paragraph{Checkerboard code.}
	The checkerboard code has one qubit at each vertex of a simple-cubic
	lattice~\cite{Vijay16}.  Each cube with lower corner $(i,j,k)$ and even
	$i+j+k$ carries one $X$-check and one $Z$-check, both acting on its
	eight vertices.

	The selected lower corners form a face-centered cubic (FCC) lattice.
	In units of the cubic-lattice spacing, let $u,v,w$ denote translations
	by $(1,1,0)$, $(1,0,1)$, and $(0,1,1)$, respectively.
	We order the two qubit orbits by their representatives $(1,1,1)$ and
	$(0,0,0)$, the opposite vertices of the reference cube anchored at the
	origin.  Each primitive cell contains two qubits and one check of
	each type.  With $h=1+u+v+w$, the complete stabilizer map is
	\begin{equation}
		\mathsf S_{\rm CB}
		=
		\begin{pmatrix}
			h^\ast&0\\
			h&0\\
			0&h^\ast\\
			0&h
		\end{pmatrix}
		=
		\begin{pmatrix}
			1+\bar u+\bar v+\bar w&0\\
			1+u+v+w&0\\
			0&1+\bar u+\bar v+\bar w\\
			0&1+u+v+w
		\end{pmatrix}.
		\label{eq:checkerboard_code_stabilizer_map}
	\end{equation}

	The checkerboard code has Type-I fracton order.  Applying Pauli-$X$
	or Pauli-$Z$ operators across a rectangular patch leaves four corner
	defects.  Although individually immobile, these defects can form
	composites that move along a line or within a plane.

	\paragraph{X-cube code.}
	The X-cube code has one qubit on each edge of a simple-cubic
	lattice~\cite{Vijay16,Song22}.  Each cube carries an $X$-check acting
	on its twelve edges, while each vertex carries three $Z$-checks,
	each acting on the four incident edges in one coordinate plane.
	The three $Z$-checks at each vertex multiply to the identity, leaving
	two independent generators.

	Let $x,y,z$ denote unit translations.  At a reference vertex, we order
	the qubit species by the positive $x,y,z$ edges and choose $(b_x,b_y)$
	as the $yz$- and $xz$-plane $Z$-checks.  With $f=1+x$, $g=1+y$,
	and $h=1+z$, the complete stabilizer map is
	\begin{equation}
		\mathsf S_{\rm XC}
		=
		\begin{pmatrix}
			0&f^\ast&0\\
			g^\ast&0&0\\
			h^\ast&h^\ast&0\\
			0&0&gh\\
			0&0&fh\\
			0&0&fg
		\end{pmatrix}
		=
		\begin{pmatrix}
			0&1+\bar x&0\\
			1+\bar y&0&0\\
			1+\bar z&1+\bar z&0\\
			0&0&(1+y)(1+z)\\
			0&0&(1+x)(1+z)\\
			0&0&(1+x)(1+y)
		\end{pmatrix}.
		\label{eq:xcube_code_stabilizer_map}
	\end{equation}

	The X-cube code has Type-I fracton order: rectangular Pauli-$Z$
	membranes create four immobile fractons at their corners, while
	straight Pauli-$X$ strings create lineon pairs at their endpoints.
	Each lineon moves only along the string direction.

	\subsection{Three-dimensional color codes}
	\label{subsec:three_dimensional_color_codes}

	The three-dimensional color code is defined on a tetrahedralization
	of the body-centered cubic (bcc) lattice~\cite{Bombin07,Butt26}.
	Each tetrahedron carries one qubit and has one vertex of each of four
	colors.  Each vertex supports one $X$-check and each edge one
	$Z$-check, acting on the qubits in the incident tetrahedra:
	$S_X(v)=\prod_{t\supset v}X_t$ and
	$S_Z(e)=\prod_{t\supset e}Z_t$.

	Let $R=\mathbb F_2[x^{\pm1},y^{\pm1},z^{\pm1}]$.  Taking the
	conventional cubic lattice constant to be unity, we let $x,y,z$
	denote translations by the displacement vectors
	\begin{equation}
		x\mapsto\tfrac12(1,1,-1),
		\qquad
		y\mapsto\tfrac12(1,-1,1),
		\qquad
		z\mapsto\tfrac12(-1,1,1).
		\label{eq:color_bcc_primitive_translations}
	\end{equation}
	It follows that $xy$, $xz$, and $yz$ are unit translations along
	the three Cartesian axes of the conventional cubic cell.  Anchoring a
	primitive parallelepiped at the reference vertex $1$, its eight
	vertices are $1,x,y,z,xy,xz,yz,xyz$.  Using the diagonal from $1$
	to $xyz$, the Freudenthal triangulation partitions this
	parallelepiped according to the $3!$ orderings of the three
	primitive-coordinate steps, yielding six translation orbits of
	tetrahedra,
	\begin{equation}
		\begin{aligned}
			q_{xy}&=(1,x,xy,xyz),
			& q_{xz}&=(1,x,xz,xyz),\\
			q_{yx}&=(1,y,xy,xyz),
			& q_{yz}&=(1,y,yz,xyz),\\
			q_{zx}&=(1,z,xz,xyz),
			& q_{zy}&=(1,z,yz,xyz).
		\end{aligned}
		\label{eq:color_tetrahedron_orbits}
	\end{equation}
	Assigning the vertex $x^iy^jz^k$ the color
	$i+j+k\pmod 4$ gives the four vertices of every tetrahedron
	distinct colors, explicitly realizing the four-coloring above.
	Each subscript specifies the first two steps along the ordered path
	from $1$ to $xyz$; the omitted primitive direction is the final
	step.  We use these orbit representatives as the ordered qubit basis
	$\mathfrak B_q=(q_{xy},q_{xz},q_{yx},q_{yz},q_{zx},q_{zy})$.
	Under lattice translations, the edges in this triangulation form
	seven orbits, represented by the ordered $Z$-check basis
	$\mathfrak B_e=(e_x,e_y,e_z,e_{xy},e_{xz},e_{yz},e_{xyz})$,
	where $e_u=(1,u)$ denotes the edge joining the reference vertex $1$ to
	$u$.  Thus, each primitive unit cell contains six qubit types, one
	vertex $X$-check type, and seven edge $Z$-check types.

	The $X$- and $Z$-type stabilizer maps are
	\begin{equation}
		A
		=
		\begin{pmatrix}
			1+\bar x+\bar x\bar y+\bar x\bar y\bar z\\
			1+\bar x+\bar x\bar z+\bar x\bar y\bar z\\
			1+\bar y+\bar x\bar y+\bar x\bar y\bar z\\
			1+\bar y+\bar y\bar z+\bar x\bar y\bar z\\
			1+\bar z+\bar x\bar z+\bar x\bar y\bar z\\
			1+\bar z+\bar y\bar z+\bar x\bar y\bar z
		\end{pmatrix},
		\qquad
		B
		=
		\begin{pmatrix}
			1&\bar x&\bar x\bar y&1&0&\bar x&1\\
			1&\bar x\bar z&\bar x&0&1&\bar x&1\\
			\bar y&1&\bar x\bar y&1&\bar y&0&1\\
			\bar y\bar z&1&\bar y&0&\bar y&1&1\\
			\bar z&\bar x\bar z&1&\bar z&1&0&1\\
			\bar y\bar z&\bar z&1&\bar z&0&1&1
		\end{pmatrix}.
		\label{eq:color_A_matrix}
	\end{equation}
	The complete stabilizer map is
	$\mathsf S_{\rm color}=\operatorname{diag}(B,A)$.

	The maps $A$ and $B$ specify the tetrahedra incident on each vertex
	and edge, respectively.  Each vertex $X$-check has weight $24$.
	The edge $Z$-checks have weight six for $e_x,e_y,e_z,e_{xyz}$ and
	weight four for $e_{xy},e_{xz},e_{yz}$.

	\subsection{Bivariate bicycle codes}
	\label{subsec:bb_codes_studied}

	Each unit cell of a bivariate bicycle code contains two qubits, one on
	a horizontal link and one on a vertical link, together with one
	$X$-check and one $Z$-check~\cite{Bravyi24}.  We can label BB codes
	with a toric layout as
	$\mathrm{BB}(\bar\alpha,\bar\beta,a,b)$~\cite{Chen25}, where
	$\bar\alpha=-\alpha$, $\bar\beta=-\beta$,
	$\alpha\geq\beta\geq0$, and $a,b\in\mathbb Z$.  With
	$f=1+x+x^{\bar\alpha}y^b$ and $g=1+y+y^{\bar\beta}x^a$,
	the complete stabilizer map is
	\begin{equation}
		\mathsf S_{\rm BB}
		=
		\begin{pmatrix}
			g^\ast&0\\
			f^\ast&0\\
			0&f\\
			0&g
		\end{pmatrix}
		=
		\begin{pmatrix}
			1+\bar y+y^{\beta}x^{-a}&0\\
			1+\bar x+x^{\alpha}y^{-b}&0\\
			0&1+x+x^{\bar\alpha}y^b\\
			0&1+y+y^{\bar\beta}x^a
		\end{pmatrix}.
		\label{eq:bb_code_stabilizer_map}
	\end{equation}

	BB codes feature symmetry-enriched topological (SET) order, in which
	lattice translations permute anyon types.  The anyon period along each
	lattice direction is the smallest positive number of unit translations
	that restores every anyon type~\cite{Chen25}.
	The families 
	$\mathbb{BB}(\bar 1,\bar 1,3,3)$ and
	$\mathbb{BB}(\bar 1,\bar 1,3,\bar 3)$ are the parent classes of the
	gross $[[144,12,12]]$ and two-gross $[[288,12,18]]$ codes,
	respectively~\cite{Bravyi24,Chen25}.
	These families have maximal logical-qubit counts $k_{\max}=16$ and
	$26$, with respective anyon periods $(12,12)$ and $(762,762)$
	along the two lattice directions~\cite{Chen25}.

\section{Threshold Results}
\label{sec:optimal_thresholds_codes_studied}

The optimal threshold is determined by the intersection of the phase
boundary with the Nishimori line.
The overall threshold of each code is the lower of its two CSS-sector
thresholds.
Self-duality of Haah's, checkerboard, and BB codes enforces identical
$X$- and $Z$-sector thresholds, whereas the overall thresholds of the
X-cube and 3D color codes are set by the $Z$ sector.

\subsection{Identification of the phase transition}
\label{subsec:hts_phase_transition_identification}

We probe the phase transition through the singularity of the specific heat.
For the models considered here, the quenched free-energy expansion in
Eq.~\eqref{eq:minimum_filling_free_energy} contains only even powers of
$v=\tanh\beta$, so we use $u=v^2$. For even $N$, truncating the series
at $v^N$ retains terms through $u^{N/2}$. At fixed $p$, we write
\begin{equation}
  \begin{aligned}
    \beta[f_{\xi_\mu}]_{\mathrm{dis}}&=\beta f_0-\Phi_p(u),\\
    \Phi_p(u)&=\sum_{i=1}^{N/2}f_i(p)u^i+O(u^{N/2+1}),
    \qquad f_i(p)=\sum_k A_{2i,k}(1-2p)^k .
  \end{aligned}
  \label{eq:hts_fixed_p_graph_series}
\end{equation}
Evaluating $f_i(p)$ at a different disorder rate gives the corresponding
thermal series without repeating the disorder average.

At fixed $p$, the specific heat per unit cell can be expressed as
\begin{equation}
  \frac{C_v(p,\beta)}{k_B}
  =-\beta^2\partial_\beta^2\!\left(\beta[f_{\xi_\mu}]_{\mathrm{dis}}\right)
  =\frac{C_{v,0}(\beta)}{k_B}+\beta^2\partial_\beta^2\Phi_p(u).
  \label{eq:hts_specific_heat_relation}
\end{equation}
Here $C_{v,0}$ is the analytic background from $\beta f_0$; neither it nor
the nonzero factor $\beta^2$ changes the singularity location at finite
positive temperature.

The derivative is evaluated term by term using
$\partial_\beta=(1-v^2)\partial_v$. Each monomial contributes
\begin{equation}
  \partial_\beta^2\!\left(f_i u^i\right)
  =f_i\left[2i(2i-1)u^{i-1}-8i^2u^i+2i(2i+1)u^{i+1}\right].
  \label{eq:hts_specific_heat_monomial_rule}
\end{equation}
The uncomputed $O(u^{N/2+1})$ remainder can contribute to the second
derivative starting at $u^{N/2}$. We therefore retain the derivative series
only through $u^{N/2-1}$, discarding the $u^{N/2}$ and $u^{N/2+1}$ terms
generated from the retained free-energy polynomial.

Let $s$ denote the lowest exponent with a nonzero coefficient in the
retained derivative series. Factoring out $u^s$ gives a series with a
nonzero constant term:
\begin{equation}
  \mathcal C_p(u)
  \coloneqq u^{-s}\partial_\beta^2\Phi_p(u)
  =\sum_{n=0}^{n_{\max}}b_n(p)u^n+O(u^{n_{\max}+1}),
  \qquad n_{\max}=N/2-1-s .
  \label{eq:hts_shifted_specific_heat_series}
\end{equation}
The auxiliary series $\mathcal C_p(u)$ retains the specific heat's
physical singularity at $0<u_c<1$.

\begin{figure}[!tp]
  \centering
  \includegraphics{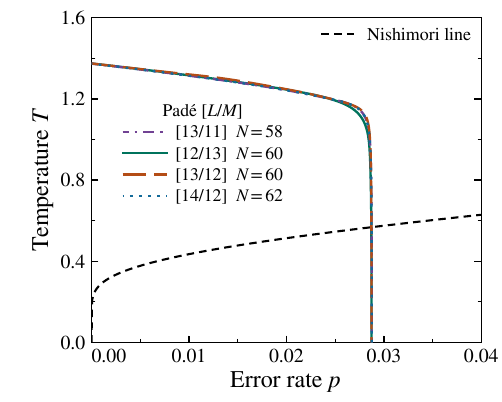}
  \caption{Phase boundary of the
  $\mathbb{BB}(\bar1,\bar1,3,3)$ code under spacetime decoherence.
  The colored curves correspond to different Dlog--Pad\'e approximants
  labeled by $[L/M]$ and $N$, while the black dashed line denotes the
  Nishimori line.
  Four approximants are shown for illustration; a larger set is used
  for threshold estimation.
  The displayed phase boundaries and their intersections with the
  Nishimori line agree closely.}
  \label{fig:hts_bb1133_phase_ranges}
\end{figure}

\begin{figure}[!tp]
  \centering
  \includegraphics{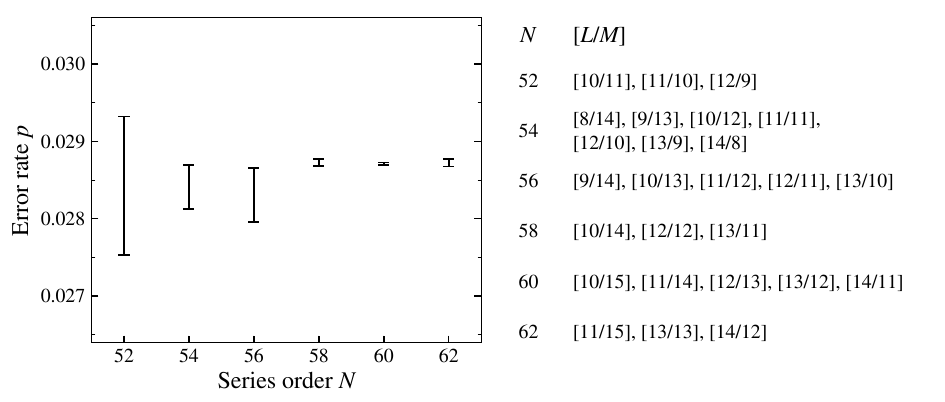}
  \caption{Order dependence and spread of the threshold estimates for the
  $\mathbb{BB}(\bar1,\bar1,3,3)$ code under spacetime decoherence.
  We use 26 Dlog--Pad\'e approximants across six expansion orders,
  $N=52,54,\ldots,62$.
  At each order, the vertical bar spans the minimum and maximum
  estimates across the different approximants.
  The clustering supports a stable threshold estimate, while the
  spread across these orders informs its conservative uncertainty.}
  \label{fig:hts_bb1133_order_ranges}
\end{figure}

\begin{figure}[htbp]
  \centering
  \includegraphics[width=0.9\textwidth]{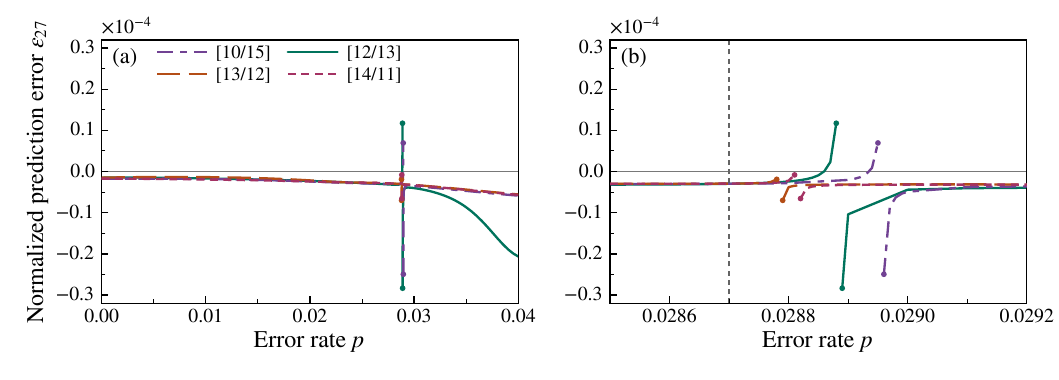}
  \caption{Prediction deviation for the
  $\mathbb{BB}(\bar1,\bar1,3,3)$ code.
  The colored curves show $\varepsilon_J$ with $J=27$
  [Eq.~\eqref{eq:hts_bb1133_holdout_residual}] for four different
  $N=60$ Dlog--Pad\'e approximants.
  The reference coefficient $b_J(p)$ is computed from the $N=62$
  expansion and excluded from the prediction input.
  (a) Full error-rate range; (b) enlargement near the threshold.
  Despite local spikes near the threshold, the plotted deviations
  remain of order $10^{-5}$ or smaller.}
  \label{fig:hts_bb1133_holdout}
\end{figure}

\subsection{Threshold Extraction}
\label{subsec:hts_threshold_analysis}

We use Dlog--Pad\'e analysis to locate the singularities of the series $\mathcal C_p(u)$.
This method is widely used to extract critical points from
finite-order series~\cite{Hellmund03,Hellmund06}.
For each $p$, we construct the $[L/M]$ Pad\'e approximant
to the logarithmic derivative of $\mathcal C_p(u)$,
\begin{equation}
  \frac{d}{du}\log\mathcal C_p(u)
  =\frac{\mathcal C_p'(u)}{\mathcal C_p(u)}
  \simeq\frac{A_L(u)}{B_M(u)},\qquad B_M(0)=1,
  \label{eq:hts_dlog_pade_approximant}
\end{equation}
where $A_L$ and $B_M$ are polynomials of degrees $L$ and $M$,
respectively. Coefficient matching determines the approximant from the first
$J=L+M+2$ coefficients $b_0(p),\ldots,b_{J-1}(p)$ of
$\mathcal C_p$, subject to $J\le N/2-s$.

We track the poles $u_\star(p)$ by matching roots of $B_M$ at
neighboring values of $p$.
For a real pole $u_\star\in(0,1)$, we use $u=\tanh^2\beta$
and $T=1/\beta$ to obtain the corresponding reduced temperature,
\begin{equation}
  T_\star=\frac{1}{\operatorname{arctanh}(\sqrt{u_\star})}.
  \label{eq:hts_dlog_pole_temperature}
\end{equation}
As $p$ varies, the temperature curve $T_\star(p)$ of each tracked
pole defines a branch in the $(p,T)$ plane. The order--disorder
boundary is traced by comparing these branches across Pad\'e approximants.

The optimal threshold $p_{\rm th}$ is determined by the intersection of this
order--disorder phase boundary with the Nishimori line, as illustrated with the $\mathbb{BB}(\bar1,\bar1,3,3)$ code under noisy measurements in Fig.~\ref{fig:hts_bb1133_phase_ranges}.
A descending branch that crosses the line from above as $p$
increases gives a direct estimate of $p_{\rm th}$.
When the relevant branches instead end in real-pole collisions
above the line, we use the collision region and its stability
across $N$ and $[L/M]$ to estimate the crossing.

At a given expansion order $N$, threshold estimates are obtained from
different $[L/M]$ approximants. The threshold value is inferred from the
clustering of those individual estimates. The smallest and largest
estimates define an interval for the inferred threshold at a fixed order.
Convergence is assessed over the highest available orders by checking for
residual drift with $N$. The reported threshold and its uncertainty range
are determined on a conservative basis, with the range covering the
spread across both $N$ and $[L/M]$ at highest orders.

As a further consistency check, we withhold the highest-order computed
coefficient $b_J(p)$ and reconstruct $\mathcal C_p(u)$ from the
lower-order coefficients $b_0(p),\ldots,b_{J-1}(p)$.
The reconstructed series is re-expanded through $u^J$ to obtain the
prediction $\widehat b_J(p)$.
Comparing the predicted coefficient $\widehat b_J(p)$ with the directly
computed $b_J(p)$ gives the deviation
\begin{equation}
  \varepsilon_J(p)
  =\frac{\widehat b_J(p)-b_J(p)}
        {|\widehat b_J(p)|+|b_J(p)|}.
  \label{eq:hts_bb1133_holdout_residual}
\end{equation}
A very small $|\varepsilon_J(p)|$ indicates that the approximants
already capture the next-order series contribution.
This predictive agreement provides further evidence for the convergence
of the threshold estimates.

We illustrate the procedure with the
$\mathbb{BB}(\bar1,\bar1,3,3)$ code under noisy syndrome measurements.
The free-energy series is computed through order $v^{62}$,
providing 28 coefficients of $\mathcal C_p(u)$.
The four representative phase-boundary branches in
Fig.~\ref{fig:hts_bb1133_phase_ranges} agree closely near their
intersections with the Nishimori line.
Figure~\ref{fig:hts_bb1133_order_ranges} compares threshold estimates
across expansion orders from $N=52$ to $N=62$.
We estimate the central threshold value from the tight clustering at
the highest three orders and its uncertainty conservatively from the
spread across all examined orders and Pad\'e approximants.
Moreover, Fig.~\ref{fig:hts_bb1133_holdout} compares predictions from
lower-order input with the additional coefficient computed at $N=62$,
showing deviations typically of order $10^{-5}$ or smaller that further
support convergence of the threshold estimates.
We thus determine its optimal threshold, $p_{\mathrm{th}}=0.029(2)$.

\begin{figure}[t]
  \centering
  \includegraphics{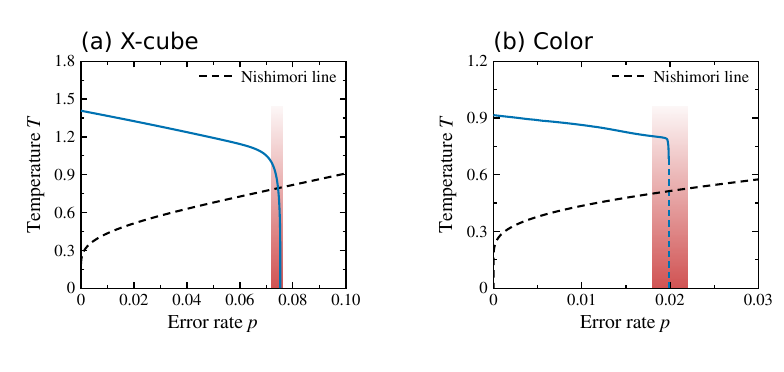}
  \caption{Optimal thresholds of (a) the X-cube code and (b) the 3D color
  code under perfect measurements.
  The displayed calculations use free-energy expansion orders $N=36$
  for the X-cube code and $N=148$ for the 3D color code.
  Blue solid curves correspond to representative $[4/7]$ and $[30/31]$
  Dlog--Pad\'e approximants, respectively, while black dashed curves
  denote the Nishimori line.
  Shaded bands indicate the threshold uncertainty ranges.
  For the X-cube code, the intersection of the phase boundary with the
  Nishimori line gives $p_{\mathrm{th}}=0.074(2)$.
  For the 3D color code, the collision point lies above the Nishimori line.
  The vertical blue dashed line connects this point to its projection
  on the $p$ axis. We estimate $p_{\mathrm{th}}=0.020(2)$ from the
  collision region.
  These values agree with Monte Carlo estimates of
  $0.075(2)$~\cite{Song22} and $0.019$~\cite{Kubica18}, respectively.}
  \label{fig:hts_color_x_cube_code_capacity}
\end{figure}

\begin{figure}[t]
  \centering
  \includegraphics{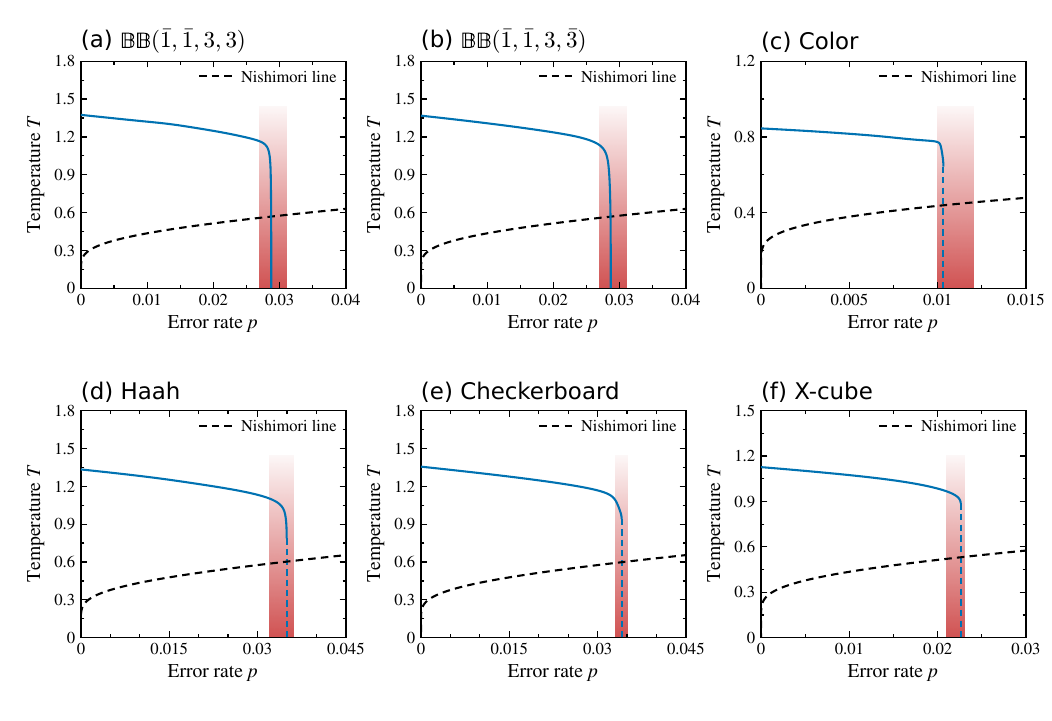}
  \caption{Optimal thresholds for six codes under spacetime decoherence,
  including the $\mathbb{BB}(\bar1,\bar1,3,3)$ code used to illustrate
  the procedure.
  The phase boundaries shown are computed using expansion orders $N=60$
  for the two BB codes, $N=178$ for the 3D color code, $N=68$ for
  Haah's code, $N=68$ for the checkerboard code, and $N=88$ for the
  X-cube code.
  Blue solid curves correspond to representative Dlog--Pad\'e approximants
  $[13/12]$ in (a,b), $[35/40]$ in (c), $[14/14]$ in (d,e), and
  $[18/18]$ in (f), while black dashed curves denote the Nishimori line.
  Vertical blue dashed lines in (c)--(f) connect the collision points
  to their projections on the $p$ axis.
  Shaded bands indicate threshold uncertainty ranges.
  The two BB codes have nearly identical optimal thresholds,
  $p_{\mathrm{th}}=0.029(2)$, despite their different maximal
  logical-qubit counts, $k_{\max}=16$ and $26$~\cite{Chen25}.
  The optimal thresholds for the 3D color, Haah's, checkerboard, and
  X-cube codes are $p_{\mathrm{th}}=0.011(1)$, $0.034(2)$, $0.034(1)$,
  and $0.022(1)$, respectively.}
  \label{fig:hts_five_model_phase_boundaries}
\end{figure}

We apply the same procedure to the benchmark cases
(Fig.~\ref{fig:hts_color_x_cube_code_capacity}) and to the other codes
considered here (Fig.~\ref{fig:hts_five_model_phase_boundaries}).
At finite expansion order, some pole branches terminate above the
Nishimori line when two real poles of a Dlog--Pad\'e approximant coalesce.
The collision points remain stable in $p$ across expansion orders
and Pad\'e approximants.

\subsection{Graph enumeration}
\label{subsec:hts_graph_enumeration}

We compute the free-energy coefficients $A_{m,k}$ by exhaustive
enumeration of connected generator supports
[Eq.~\eqref{eq:minimum_filling_series_coefficients}], with the achieved
expansion orders and enumeration counts summarized in
Table~\ref{tab:hts_graph_enumeration_counts}.

\begin{table}[tbp]
  \caption{Expansion orders and graph-enumeration counts for the codes
  studied. Here $N$ is the highest computed free-energy expansion order
  in $v$. The counts include only connected generator configurations
  and their isomorphism classes with nonzero disorder-averaged
  contributions through $v^N$. Configurations are counted up to lattice
  translation; isomorphism classes group configurations with the same
  generator--interaction incidence structure. Evaluating one
  representative per class substantially reduces the number of weight
  calculations.}
  \label{tab:hts_graph_enumeration_counts}
  \centering
  \renewcommand{\arraystretch}{1.15}
  \begin{tabular}{lc@{\hspace{\dimexpr2\tabcolsep+0.8em\relax}}rr}
    \toprule
    Code & $N$ & \shortstack{Connected\\generator configurations}
         & \shortstack{Graph-isomorphism\\classes}\\
    \midrule
    \multicolumn{4}{c}{\textit{Perfect syndrome measurements}}\\
    \addlinespace[2pt]
    X-cube & 56 & 65,669,975 & 520,990 \\
    3$D$ color & 148 & 16,904,887,502 & 31,804,806 \\
    \midrule
    \multicolumn{4}{c}{\textit{Noisy syndrome measurements}}\\
    \addlinespace[2pt]
    $\mathbb{BB}(\bar 1,\bar 1,3,3)$ & 62 & 42,434,278,564 & 7,365,948 \\
    $\mathbb{BB}(\bar 1,\bar 1,3,\bar 3)$ & 62 & 48,093,311,901 & 1,280,951 \\
    3$D$ color & 178 & 93,653,204,867 & 83,972,838 \\
    Haah & 68 & 18,874,266,699 & 51,387,689 \\
    Checkerboard & 72 & 170,301,947,729 & 12,795,052 \\
    X-cube & 90 & 17,430,370,903 & 13,286,266 \\
    \bottomrule
  \end{tabular}
\end{table}

For the sectors of the codes studied here, each admissible finite-support graph
has a unique filling. Each calculation uses one elementary graph
generator per primitive cell: the complementary stabilizer map
$\partial_1^\nu$ is $B$ for the $X$ sectors of the two BB codes,
Haah's code, and the checkerboard code, and $A$ for the $Z$ sectors
of the 3D color and X-cube codes. These maps are single nonzero
columns over the integral domain $R_{\rm s}$ and are therefore
injective on finite-support configurations
(Sec.~\ref{sec:quantum_codes_studied}). The spacetime graph-generator
maps retain this injectivity through their spatial components
[Eq.~\eqref{eq:higher_cross_sector_correspondence}].
Spatial and spacetime exactness
[Eqs.~\eqref{eq:static_middle_exactness} and
\eqref{eq:graph_completeness}] ensure that these fillings exist on
the infinite lattice. Their uniqueness then gives $\rho(\gamma)=1$
by Eq.~\eqref{eq:global_filling_probability}.

Only connected generator supports contribute to the series coefficients.
Let $g_c$ denote the set of interaction locations in the elementary
boundary of a generator placement $c$. Two placements $c$ and $d$
are adjacent if $g_c\cap g_d\ne\varnothing$. A support $\gamma$ is
disconnected precisely when it admits a partition
$\gamma=\gamma_1\sqcup\gamma_2$ into two nonempty parts such that
$\left(\bigcup_{c\in\gamma_1}g_c\right)\cap
\left(\bigcup_{d\in\gamma_2}g_d\right)=\varnothing$.
These unions retain interaction locations that cancel modulo two
in the boundary of the full configuration.
Each subfilling splits uniquely into two pieces whose boundaries lie
in the two disjoint unions above, so their graph factors multiply at
fixed disorder. With $\rho=1$,
Eq.~\eqref{eq:minimum_filling_local_polynomial} therefore gives
$\mathcal P(\gamma'_1\sqcup\gamma'_2)
=\mathcal P(\gamma'_1)\mathcal P(\gamma'_2)$
for any $\gamma'_i\subseteq\gamma_i$.
Taking the logarithm and applying
Eq.~\eqref{eq:minimum_filling_mobius_weight} gives
$\mathcal W(\gamma)=0$ for every disconnected support $\gamma$
and every disorder realization.

The cover expansion in Eq.~\eqref{eq:minimum_filling_disorder_weight},
with $\rho=1$, determines the finite-order contributions of each
connected support $\gamma$. It sums over all multisets of nonempty
subfillings whose union
is $\gamma$; these subfillings may overlap, repeat, or be disconnected.
A cover $\mathcal A$ contributes at degree
$m(\mathcal A)=\sum_i|g_i|$, where $g_i$ is the boundary of its
$i$th subfilling. Thus a support can contribute through $v^N$ only
if it admits a cover with $m(\mathcal A)\le N$.
The cover consisting of $\gamma$ alone has degree $|g|$, where $g$
is the boundary of $\gamma$, whereas overlapping subfillings can
cover the same support at a lower total degree. Cancellations
between cover contributions can also raise the first nonzero order
of the disorder-averaged weight above $|g|$.

We enumerate connected generator supports up to translation by adapting
Redelmeier's method for fixed polyominoes~\cite{Redelmeier81} to the
graph of generator placements defined above.
We order placements lexicographically in primitive lattice
coordinates and translate the least placement of each support
to the origin. This selects a unique representative of each
translation class.
Starting with the generator at the origin, we grow a support $\gamma$
by trying its eligible neighbors in turn. These are unoccupied
placements that follow the origin in the chosen order, are adjacent
to at least one generator in $\gamma$, and have not been excluded.
The branch that adds a placement $c$ inherits the remaining candidates
and exclusions and gains any new eligible neighbors of $c$. On
returning to $\gamma$, we restore its search state and exclude $c$
from all later sibling branches and their descendants.

For any finite bound on generator count, this search visits each
connected support within that bound exactly once up to translation.
At the first divergence of two search paths, the earlier branch
retains its chosen placement in every descendant, whereas the
later branch excludes it. The same support therefore cannot be
generated by both paths.
To reach a connected target support $\Gamma$ rooted as above,
choose at each step the first candidate belonging to $\Gamma$.
Every earlier candidate lies outside $\Gamma$, so excluding it
does not prevent reaching the target. Connectivity guarantees
an eligible neighbor in $\Gamma\setminus\gamma$ until
$\gamma=\Gamma$.

For each model at its target order $N$, we used the lattice geometry
to derive a sufficient storage cutoff $B_{\rm save}(N)$: every
support admitting a cover with $m(\mathcal A)\le N$ has
$|g|\le B_{\rm save}(N)$. We therefore restrict graph-isomorphism
tests to supports within this bound.
During growth, we allow the boundary size to exceed $B_{\rm save}(N)$,
since adding a generator can cancel locations on the current
boundary and bring its size back within the bound.
The supporting data for each model are provided
in the accompanying Zenodo repository~\cite{Wang2026Zenodo}.

For each code and target order $N$, we numerically verify that the
restricted search includes all contributions through $v^N$.
We repeated the enumeration with
progressively looser restrictions on boundary growth and obtained
unchanged coefficients $A_{m,k}$ for $m\le N$.
Separate enumerations for the same model at different target orders,
with the corresponding storage cutoffs, also yielded identical
coefficients over their common range of orders.

We group retained supports by isomorphism of their generator--interaction
incidence graphs. For each support, we form this bipartite graph by
joining each generator $c$ to the interaction locations in $g_c$.
Two such graphs are isomorphic if they coincide after relabeling
vertices within each of the two types.
Supports with isomorphic incidence graphs have the same
disorder-averaged weight $\mathcal W_{\mathrm{av}}$ when the couplings
have a common magnitude and independent, identically distributed signs.
The isomorphism preserves all subfilling boundaries up to relabeling,
so corresponding covers contribute equally after disorder averaging.

We evaluate this weight for one representative $\gamma_a$
of each class and obtain the coefficients per unit cell as
\begin{equation}
  A_{m,k}=\sum_a M_a\,[q_{\mathrm{dis}}^k v^m]\,
  \mathcal W_{\mathrm{av}}(\gamma_a;q_{\mathrm{dis}},v),
  \qquad m\le N,
  \label{eq:hts_class_coefficient_assembly}
\end{equation}
where the embedding multiplicity $M_a$ counts the distinct
generator supports in class $a$ up to lattice translation.
The operator $[q_{\mathrm{dis}}^k v^m]$ extracts the coefficient of
$q_{\mathrm{dis}}^k v^m$ from the representative weight.
In Table~\ref{tab:hts_graph_enumeration_counts}, we include a class $a$
if the coefficient of $v^m$ in its
disorder-averaged weight is a nonzero polynomial in
$q_{\mathrm{dis}}$ for some $m\le N$.
Each contributing isomorphism class adds $M_a$ to the configuration
count and one to the class count.

\section{Gauge-invariant order parameters}
\label{sec:code_induced_gauge_theories}

We construct generalized Wilson observables using the gauge structure
of Sec.~\ref{subsec:SI_gauge_model} and the code matrices and bases of
Sec.~\ref{sec:quantum_codes_studied}, suppressing model subscripts on
$A,B$.  We retain $R_{\rm st}=R_{\rm s}[t^{\pm1}]$, with $t$ translating
by one measurement round, and write $\delta_t\coloneqq t-1$ and
$\bar\delta_t=t^{-1}-1$; bars on Laurent polynomials denote inversion.
For a translation $\lambda$ and integer $n\geq1$, set
\begin{equation}
 [n]_\lambda\coloneqq1+\lambda+\cdots+\lambda^{n-1}.
 \label{eq:translation_sum}
\end{equation}
For a spacetime spin support $\Sigma\in F_1^\mu$, write its gauge-charge pattern as
$\mathfrak q_\mu[\Sigma]\coloneqq(\Square_2^\mu)^\dagger\Sigma$.
We use $p\in R_{\rm s}^q$ for a spatial spin support, embedded at round
$0$ as $\Sigma=(0,p)$. Its spatial gauge-charge pattern is $A^\dagger p$
in the $X$-sector and $B^\dagger p$ in the $Z$-sector.

\subsection{General construction}
\label{subsec:SI_generalized_wilson}
\label{subsec:general_construction}

A spin monomial $W_\Sigma$ is gauge invariant precisely when
$\mathfrak q_\mu[\Sigma]=0$ [Eq.~\eqref{eq:SI_wilson_neutrality}].
On the infinite lattice, Eq.~\eqref{eq:dagger_side_exactness} gives
every finite-support solution in the form
$\Sigma=\widetilde{\Square}_1^\nu g=(\Square_1^\mu)^\dagger g$,
with finite-support $g\in\widetilde F_1^\nu\cong F_0^\mu$ and
complementary sector $\nu$.  Thus $W_\Sigma$ is a product of
interaction monomials, giving the discrete Stokes relation
\begin{equation}
 W_\Sigma(s)
 =\prod_\ell W_{(\Square_1^\mu)^\dagger\ell}(s)^{g_\ell}.
 \label{eq:SI_discrete_stokes}
\end{equation}
By Eq.~\eqref{eq:graph_completeness}, two choices $g,g'$ give the same
support precisely when $g'-g=\widetilde{\Square}_2^\nu h$ for some
$h\in\widetilde F_2^\nu$.
The quenched coupling signs are not part of these spin monomials.

For $\Sigma_X=(u,v)\in F_1^X$, the gauge-invariance condition reads
\begin{equation}
 \mathfrak q_X[\Sigma_X]
 =\begin{pmatrix}(\partial_2^X)^\dagger u\\
                  \bar\delta_t u+A^\dagger v\end{pmatrix}=0.
 \label{eq:SI_wilson_charge}
\end{equation}
The two components impose the static stabilizer-relation constraint
and temporal closure.

For a spatial support $p\in R_{\rm s}^q$ with $X$-check syndrome
$A^\dagger p$, temporal extension over $n\geq1$ rounds gives
\begin{equation}
 \Sigma_X(p;n)
 =\begin{pmatrix}t[n]_t A^\dagger p\\(1+t^n)p\end{pmatrix}
 =(\Square_1^X)^\dagger
   \begin{pmatrix}t[n]_t p\\0\end{pmatrix}.
 \label{eq:SI_wilson_suspension}
\end{equation}
Here $\bar\delta_t\,t[n]_t=1+t^n$ over $\mathbb F_2$, and gauge
invariance follows from $\Square_1^X\Square_2^X=0$.
Exchanging $X,A$ with $Z,B$ gives the complementary construction.

Copies of $p$ at rounds $0,n$ create and annihilate the syndrome
$A^\dagger p$; the upper component supplies its history at rounds
$1,\ldots,n$.  A string creating endpoint anyons therefore yields
loop-type spacetime support, while a membrane creating a boundary
loop of excitations yields surface-type support.  The same formula
uses the actual excitation patterns of wireframe and fractal supports,
without requiring a smooth boundary or spatial mobility of individual
excitations.  Varying $p$ and $n$ to obtain growing supports gives
generalized Wilson families.

Generalized Wilson observables can also be supported within a single
time slice.  Such purely spatial supports arise by combining spatial
support vectors whose excitation patterns cancel, as in the loop, wireframe,
and tube constructions of
Secs.~\ref{subsec:color_gauge_supports}--\ref{subsec:checkerboard_gauge_supports}.

The cohomological degree shift in
Eq.~\eqref{eq:SI_dual_homology_preservation} connects this construction
to the $\operatorname{Ext}$ classification of code excitation
sectors~\cite{RubaYang24,Song26}.
These sector labels can organize generalized Wilson families; their
thermodynamic scaling determines whether they distinguish phases.

	\subsection{Bivariate bicycle codes}
	\label{subsec:bb_gauge_supports}
	
	BB codes are self-dual, so we study the \(X\)-sector.  For
	\(\mathrm{BB}(\bar\alpha,\bar\beta,a,b)\), with
	\(\bar\alpha=-\alpha\) and \(\bar\beta=-\beta\), the check maps are
	\(A=(f,g)^T\) and \(B=(\bar g,\bar f)^T\), where
	\(f=1+x+x^{\bar\alpha}y^b\) and
	\(g=1+y+y^{\bar\beta}x^a\).
	
	Auxiliary spins occupy the links of a cubic spacetime lattice with two
	spatial directions and one temporal direction.  The interaction map is
	\begin{equation}
		(\Square_1^X)^\dagger
		=
		\begin{pmatrix}
			\bar f & \bar g & 0\\
			\bar\delta_t & 0 & \bar g\\
			0 & \bar\delta_t & \bar f
		\end{pmatrix}.
		\label{eq:bb_interaction_incidence_map}
	\end{equation}
	The first two columns combine the three-site support of \(\bar f\) or
	\(\bar g\) with \(\bar\delta_t\): each gives a five-body interaction
	between adjacent time slices, with one non-nearest-neighbor spatial term.
	The third combines \(\bar g\) and \(\bar f\) in one slice, giving a
	six-body interaction with two non-nearest-neighbor spatial terms.

	Since \(\ker A=\ker B=0\), we take \(C_2^X=C_2^Z=0\).
	The gauge map \(\Square_2^X\) and its adjoint are therefore
	\begin{equation}
		\Square_2^X
		=
		\begin{pmatrix}
			\delta_t\\
			f\\
			g
		\end{pmatrix},
		\qquad
		(\Square_2^X)^\dagger
		=
		\begin{pmatrix}
			\bar\delta_t&\bar f&\bar g
		\end{pmatrix}.
		\label{eq:bb_gauge_maps}
	\end{equation}
	A gauge-invariant spin support \(\Sigma_X=(u,v,w)^T\) satisfies
	\begin{equation}
		\mathfrak q_X[\Sigma_X]
		=
		(\Square_2^X)^\dagger\Sigma_X
		=
		\bar\delta_t u+\bar f v+\bar g w
		=
		0.
		\label{eq:bb_gauge_charge_pattern}
	\end{equation}
	
	The ideal of charge patterns generated by \((\Square_2^X)^\dagger\) is
	\begin{equation}
		\mathcal I_{\rm st}
		\coloneqq
		\operatorname{im}(\Square_2^X)^\dagger
		=
		(\bar\delta_t,\bar f,\bar g)
		\subset R_{\rm st}.
		\label{eq:bb_spacetime_charge_ideal}
	\end{equation}
	
	For \(\mathrm{BB}(\bar 1,\bar 1,3,3)\), we lift the Laurent ideal
	\(\mathcal I_{\rm st}\) to the ordinary polynomial ring
	\(S=\mathbb F_2[\bar t,t,\bar y,\bar x,y,x]\) to compute a reduced
	Gr\"obner basis.  Here the barred variables are independent
	indeterminates.  The quotient map
	\(\pi:S\twoheadrightarrow R_{\rm st}\) leaves \(t,x,y\) unchanged and
	sends \((\bar t,\bar x,\bar y)\) to
	\((t^{-1},x^{-1},y^{-1})\).  The lifted ideal is
	\begin{equation}
		\begin{aligned}
			\mathcal J_{\rm st}
			\coloneqq\pi^{-1}(\mathcal I_{\rm st})
			=\bigl(&1+\bar t,\,1+t\bar t,\,
			1+\bar x+x\bar y^3,\\[-2pt]
			&1+\bar y+\bar x^3y,\,1+x\bar x,\,
			1+y\bar y\bigr)\subset S.
		\end{aligned}
		\label{eq:bb_spacetime_polynomial_lift}
	\end{equation}
	With the lexicographic order
	\(\bar t>t>\bar y>\bar x>y>x\), the reduced Gr\"obner basis of
	\(\mathcal J_{\rm st}\) is
	\begin{equation}
		\begin{aligned}
			\mathcal G_{\rm st}
			&=\{h_1,\ldots,h_7\},\\
			h_1&=1+\bar t,\\
			h_2&=1+t,\\
			h_3&=1+x+x^3+x^5+x^6,\\
			h_4&=1+x^3+y+xy+x^2y,\\
			h_5&=1+x+x^4+y+y^2,\\
			h_6&=1+x^2+x^4+x^5+\bar x,\\
			h_7&=1+x+x^2+x^3+y+\bar y .
		\end{aligned}
		\label{eq:bb_spacetime_groebner_basis}
	\end{equation}
	Their images under \(\pi\), denoted by the same symbols \(h_i\), generate
	\(\mathcal I_{\rm st}\).

	Choose finite spin supports \(\Sigma_i\) with
	\(\mathfrak q_X[\Sigma_i]=h_i\).  Every allowed charge move
	\(\Delta q\in\mathcal I_{\rm st}\) is realized by a finite
	\(R_{\rm st}\)-linear combination:
	\begin{equation}
		\mathfrak q_X\!\left[\sum_{i=1}^7\alpha_i\Sigma_i\right]
		=
		\sum_{i=1}^7\alpha_i h_i
		=
		\Delta q ,
		\qquad
		\alpha_i\in R_{\rm st}.
		\label{eq:bb_charge_move_realization}
	\end{equation}

	For example, \(h_3\) describes the charge move
	\(1\longmapsto1+h_3=x+x^3+x^5+x^6\) along the \(x\) direction.
	One realization is \((0,p_3)\), with
	\(\mathfrak q_X[(0,p_3)]=A^\dagger p_3=h_3\), where
	\begin{equation}
		p_3
		\coloneqq
		\begin{pmatrix}
			x^2y+x^3y+x^5\\
			x^3y^{-1}+x^4y^{-1}+x^4+x^6y^{-2}+x^6y^{-1}+x^6
		\end{pmatrix}.
		\label{eq:bb_h3_spin_support}
	\end{equation}
	The support spans lattice coordinates \(-2\) to \(1\) in the \(y\)
	direction, so \((0,p_3)\) is an open charged ribbon of finite width.
	Combining translated supports
	\(\Sigma_i\) realizes the charge moves above.  Two allowed transport
	supports with the same endpoint charge patterns have a gauge-invariant
	sum.  Choosing them along paths that stay a bounded distance apart gives
	closed ribbons whose length can grow while their width remains finite.

	\subsection{Three-dimensional color code}
	\label{subsec:color_gauge_supports}
	
	The three-dimensional color code induces two distinct gauge theories on
	the four-dimensional spacetime extension of the tetrahedral lattice.
	
	We first consider the \(Z\)-sector.  Its interaction map is
	\begin{equation}
		(\Square_1^Z)^\dagger
		=
		\left(
		\begin{array}{cccccc|c}
			1&1&y&yz&z&yz&0\\
			x&xz&1&1&xz&z&0\\
			xy&x&xy&y&1&1&0\\
			1&0&1&0&z&z&0\\
			0&1&y&y&1&0&0\\
			x&x&0&1&0&1&0\\
			1&1&1&1&1&1&0\\ \hline
			\bar\delta_t&0&0&0&0&0&1+\bar x+\bar x\bar y+\bar x\bar y\bar z\\
			0&\bar\delta_t&0&0&0&0&1+\bar x+\bar x\bar z+\bar x\bar y\bar z\\
			0&0&\bar\delta_t&0&0&0&1+\bar y+\bar x\bar y+\bar x\bar y\bar z\\
			0&0&0&\bar\delta_t&0&0&1+\bar y+\bar y\bar z+\bar x\bar y\bar z\\
			0&0&0&0&\bar\delta_t&0&1+\bar z+\bar x\bar z+\bar x\bar y\bar z\\
			0&0&0&0&0&\bar\delta_t&1+\bar z+\bar y\bar z+\bar x\bar y\bar z
		\end{array}
		\right).
		\label{eq:color_z_interaction_map}
	\end{equation}
	The relations among the \(B\) checks are generated by the injective map
	\(\partial_2^Z\):
	\begin{equation}
		\partial_2^Z
		=
		\begin{pmatrix}
			1+\bar x&\bar x\\
			1+\bar y&\bar y\\
			1+\bar z&\bar z\\
			0&1+\bar x\bar y\\
			0&1+\bar x\bar z\\
			0&1+\bar y\bar z\\
			1+\bar x\bar y\bar z&1
		\end{pmatrix}
		:
		R_{\rm st}^2\longrightarrow R_{\rm st}^7 .
		\label{eq:color_B_relation_map}
	\end{equation}
	The map for gauge transformations is
	\begin{equation}
		\Square_2^Z
		=
		\begin{pmatrix}
			\partial_2^Z&\delta_t I_7\\
			0&B
		\end{pmatrix}.
		\label{eq:color_z_gauge_map}
	\end{equation}
	
	A finite spin support
	\(\Sigma_Z=(u,v)\in R_{\rm st}^7\oplus R_{\rm st}^6\) has charge
	pattern
	\begin{equation}
		\begin{pmatrix}
			q_1\\
			q_2
		\end{pmatrix}
		\coloneqq
		\mathfrak q_Z[\Sigma_Z]
		=
		(\Square_2^Z)^\dagger\Sigma_Z
		=
		\begin{pmatrix}
			(\partial_2^Z)^\dagger u\\
			\bar\delta_t u+B^\dagger v
		\end{pmatrix}.
		\label{eq:color_z_charge_pattern}
	\end{equation}
	Since \(\partial_2^Z\) is injective, we take \(C_3^Z=0\).
	The gauge resolution then terminates with \(\Square_3^Z
	=\binom{\delta_t I_2}{\partial_2^Z}\), whose adjoint imposes the higher
	constraint
	\begin{equation}
		(\Square_3^Z)^\dagger
		\begin{pmatrix}
			q_1\\
			q_2
		\end{pmatrix}
		=
		\bar\delta_t q_1
		+
		(\partial_2^Z)^\dagger q_2
		=
		0 .
		\label{eq:color_line_charge_constraint}
	\end{equation}
	
	For a support \(\Sigma_Z=(0,v)\) confined to one time slice,
	\(q_1=0\) and \(q_2=B^\dagger v\), so
	Eq.~\eqref{eq:color_line_charge_constraint} reduces to
	\((\partial_2^Z)^\dagger q_2=0\).

	We choose a spatial support \(p_1^{xy,xz}\in R_{\rm s}^6\),
	with spacetime embedding \((0,p_1^{xy,xz})\), where
	\begin{equation}
		\begin{aligned}
			p_1^{xy,xz}
			={}&
			\begin{pmatrix}
				1\\
				1\\
				0\\
				x\\
				0\\
				x
			\end{pmatrix},
			\\
			B^\dagger p_1^{xy,xz}
			={}&
			(1+xz)e_{xy}+(1+xy)e_{xz}.
		\end{aligned}
		\label{eq:color_elementary_membrane_support}
	\end{equation}
	The four tetrahedra \(q_{xy},q_{xz},xq_{yz},xq_{zy}\) share the edge
	\((x,xyz)\) and form the octahedron illustrated in the main text.
	This membrane patch has a rectangular charge boundary: its two
	\(e_{xy}\) edges are separated by \(xz\), and its two \(e_{xz}\)
	edges by \(xy\).

	The \(L\times M\) family of spatial supports is
	\(p_{L,M}^{xy,xz}\coloneqq
	[L]_{xy}[M]_{xz}p_1^{xy,xz}\).  Its charge is
	\(q_{L,M}^{xy,xz}\coloneqq B^\dagger p_{L,M}^{xy,xz}
	=\bigl(1+(xz)^M\bigr)[L]_{xy}e_{xy}
	+\bigl(1+(xy)^L\bigr)[M]_{xz}e_{xz}\).
	Its charge is confined to the one-dimensional boundary and satisfies
	\((\partial_2^Z)^\dagger q_{L,M}^{xy,xz}=0\), as required by
	Eq.~\eqref{eq:color_line_charge_constraint}.

	In Eq.~\eqref{eq:color_z_charge_pattern}, \(B^\dagger v\) deforms a
	line charge within a slice, while \(\bar\delta_t u\) propagates a closed
	line satisfying \((\partial_2^Z)^\dagger u=0\) between adjacent slices.
	Applying the temporal construction of Eq.~\eqref{eq:SI_wilson_suspension}
	in the \(Z\)-sector to \(p_{L,M}^{xy,xz}\) gives, for \(T\geq1\),
	\begin{equation}
		\Sigma_Z^{xy,xz}(L,M;T)
		\coloneqq
		\left(
			t[T]_{t}q_{L,M}^{xy,xz},
			(1+t^T)p_{L,M}^{xy,xz}
		\right)
		\in
		R_{\rm st}^7\oplus R_{\rm st}^6 .
		\label{eq:color_spacetime_closed_membrane_support}
	\end{equation}
	Equation~\eqref{eq:color_z_charge_pattern} gives
	\(\mathfrak q_Z[\Sigma_Z^{xy,xz}(L,M;T)]=0\).

	The two copies of \(p_{L,M}^{xy,xz}\) form the caps, joined by the
	temporal side wall \(t[T]_tq_{L,M}^{xy,xz}\).  Together they form a
	closed two-dimensional surface in spacetime, as in a two-form gauge
	theory.  Varying \(L,M,T\) gives gauge-invariant membranes of arbitrary size.

	In the \(X\)-sector, the interaction map is
	\begin{equation}
		(\Square_1^X)^\dagger
		=
		\begin{pmatrix}
			A^\dagger&0\\
			\bar\delta_t I_6&B
		\end{pmatrix}.
		\label{eq:color_x_interaction_map}
	\end{equation}
	Since \(\ker A=0\), we take \(C_2^X=0\), giving the gauge map
	\begin{equation}
		\Square_2^X
		=
		\begin{pmatrix}
			\delta_t\\
			A
		\end{pmatrix}.
		\label{eq:color_x_gauge_map}
	\end{equation}
	For \(\Sigma_X=(u,v)\in R_{\rm st}\oplus R_{\rm st}^6\), the charge is
	\(\mathfrak q_X[\Sigma_X]=\bar\delta_t u+A^\dagger v\in R_{\rm st}\),
	with no higher adjoint constraint.  Each Laurent monomial in
	\(R_{\rm st}\) represents a point charge.  For example, let
	\begin{equation}
		\rho_0=\bar x y,
		\qquad
		p_{\rho_0}
		=
		\bar x\bigl(1,\;0,\;1,\;0,\;0,\;0\bigr)^T .
		\label{eq:color_point_charge_transport_support}
	\end{equation}
	The tetrahedra \(\bar x q_{xy}\) and \(\bar x q_{yx}\) share a triangular
	face and form the bipyramid illustrated in the main text.
	The identity \(A^\dagger p_{\rho_0}=1+\rho_0\) shows that
	\((0,p_{\rho_0})\) transports a charge by \(\rho_0\).  Concatenating
	translated copies gives strings of arbitrary length and fixed transverse
	extent.  Applying Eq.~\eqref{eq:SI_wilson_suspension} to
	\([N]_{\rho_0}p_{\rho_0}\), for \(N\geq1\), gives the spacetime loop
	illustrated in the main text.  More generally, every spatial translation \(\rho\) preserving
	vertex color admits a finite spatial support \(p_\rho\in R_{\rm s}^6\) with
	\(A^\dagger p_\rho=1+\rho\).  Combining these supports along closed
	paths preserving vertex color cancels the endpoint charges, yielding
	gauge-invariant loops of arbitrarily large extent.
	
	\subsection{X-cube code}
	\label{subsec:xcube_gauge_supports}

	The X-cube code induces distinct \(Z\)- and \(X\)-sector gauge theories
	on the four-dimensional spacetime extension of the simple-cubic lattice.

	We first consider the \(Z\)-sector.  Its interaction map is
	\begin{equation}
		(\Square_1^Z)^\dagger
		=
		\begin{pmatrix}
			0&1+y&1+z&0\\
			1+x&0&1+z&0\\
			\bar\delta_t&0&0&(1+y)(1+z)\\
			0&\bar\delta_t&0&(1+x)(1+z)\\
			0&0&\bar\delta_t&(1+x)(1+y)
		\end{pmatrix}.
		\label{eq:xcube_z_interaction_map}
	\end{equation}
	The X-cube map \(B\) is injective, so we take \(C_2^Z=0\).
	The gauge map and its adjoint are
	\begin{equation}
		\Square_2^Z
		=
		\begin{pmatrix}
			\delta_t I_2\\
			B
		\end{pmatrix},
		\qquad
		(\Square_2^Z)^\dagger
		=
		\begin{pmatrix}
			\bar\delta_t I_2&B^\dagger
		\end{pmatrix}.
		\label{eq:xcube_z_gauge_maps}
	\end{equation}
	A gauge-invariant spin support
	\(\Sigma_Z=(u,v)\in R_{\rm st}^2\oplus R_{\rm st}^3\) satisfies
	\(\mathfrak q_Z[\Sigma_Z]=\bar\delta_t u+B^\dagger v=0\).
	For a spatial support \(p\in R_{\rm s}^3\), its embedding \((0,p)\) has
	charge \(B^\dagger p\), while \(\bar\delta_t u\) connects charge
	patterns on adjacent time slices.

	For \(L\geq1\), define the three string supports
	\begin{equation}
		\Sigma_x(L)\coloneqq(0,[L]_x e_x),
		\qquad
		\Sigma_y(L)\coloneqq(0,[L]_y e_y),
		\qquad
		\Sigma_z(L)\coloneqq(0,[L]_z e_z) .
		\label{eq:xcube_directional_string_supports}
	\end{equation}
	Their charge patterns are
	\begin{equation}
		\begin{aligned}
			\mathfrak q_Z[\Sigma_x(L)]
			&=
			B^\dagger [L]_x e_x
			=
			\begin{pmatrix}
				0\\
				1+x^L
			\end{pmatrix},\\
			\mathfrak q_Z[\Sigma_y(L)]
			&=
			B^\dagger [L]_y e_y
			=
			\begin{pmatrix}
				1+y^L\\
				0
			\end{pmatrix},\\
			\mathfrak q_Z[\Sigma_z(L)]
			&=
			B^\dagger [L]_z e_z
			=
			\begin{pmatrix}
				1+z^L\\
				1+z^L
			\end{pmatrix}.
		\end{aligned}
		\label{eq:xcube_directional_string_charges}
	\end{equation}
	In the independent \((b_x,b_y)\) basis, the endpoint vectors
	\((0,1)^T\), \((1,0)^T\), and \((1,1)^T\) are the three lineon types,
	restricted to motion along \(x\), \(y\), and \(z\), respectively.
	Each string creates a pair separated by \(L\).

	We join translated copies of the three strings along the edges of a cuboid.
	For \(L,M,N\geq1\), set
	\(Q_{L,M,N}\coloneqq[L]_x[M]_y[N]_z\) and define
	\begin{equation}
		\begin{aligned}
			\Sigma_{\rm wire}(L,M,N)
			&\coloneqq
			\left(0,A Q_{L,M,N}\right)\\
			&=
			\left(
			0,
			\begin{pmatrix}
				[L]_x(1+y^M)(1+z^N)\\
				[M]_y(1+x^L)(1+z^N)\\
				[N]_z(1+x^L)(1+y^M)
			\end{pmatrix}
			\right).
		\end{aligned}
		\label{eq:xcube_wireframe_family}
	\end{equation}
	Each component occupies the four cuboid edges parallel to its axis.
	At every corner, \((0,1)^T+(1,0)^T+(1,1)^T=0\), so the charges
	cancel.  Thus \(\Sigma_{\rm wire}\) is a gauge-invariant wireframe
	supported only on the cuboid edges.

	We next consider the \(X\)-sector.  Its interaction map is
	\begin{equation}
		(\Square_1^X)^\dagger
		=
		\begin{pmatrix}
			A^\dagger&0\\
			\bar\delta_t I_3&B
		\end{pmatrix}.
		\label{eq:xcube_x_interaction_map}
	\end{equation}
	The X-cube map \(A\) is injective, so we take \(C_2^X=0\).
	The gauge map and its adjoint are
	\begin{equation}
		\Square_2^X
		=
		\begin{pmatrix}
			\delta_t\\
			A
		\end{pmatrix},
		\qquad
		(\Square_2^X)^\dagger
		=
		\begin{pmatrix}
			\bar\delta_t&A^\dagger
		\end{pmatrix}.
		\label{eq:xcube_x_gauge_maps}
	\end{equation}
	A gauge-invariant spin support
	\(\Sigma_X=(u,v)\in R_{\rm st}\oplus R_{\rm st}^3\) satisfies
	\(\mathfrak q_X[\Sigma_X]=\bar\delta_t u+A^\dagger v=0\).
	For a spatial support \(p\in R_{\rm s}^3\), gauge invariance of
	\((0,p)\) reduces to \(A^\dagger p=0\).

	An open membrane carries point charges at its four corners.  For
	\(L,M\geq1\), consider the \(L\times M\) support in the
	\(xz\) plane
	\begin{equation}
		\begin{aligned}
			\Sigma_{xz}(L,M)
			&\coloneqq
			\left(0,xz[L]_x[M]_z e_y\right),\\
			\mathfrak q_X[\Sigma_{xz}(L,M)]
			&=
			A^\dagger xz[L]_x[M]_z e_y\\
			&=
			(1+x^L)(1+z^M).
		\end{aligned}
		\label{eq:xcube_open_membrane_charge}
	\end{equation}
	The image of \(A^\dagger\) contains no binomial \(1+\rho\) for a nontrivial
	spatial translation \(\rho\).  Thus no finitely supported spatial operator transports
	an isolated point charge to a distinct spatial site.

	We form a rectangular tube from four membrane faces and choose its open ends
	to be normal to the \(x\) axis.  Using the same
	\(Q_{L,M,N}=[L]_x[M]_y[N]_z\), let
	\(\Sigma_{xy}(L,M)\coloneqq(0,xy[L]_x[M]_y e_z)\) be an \(L\times M\)
	membrane in the \(xy\) plane.  The spin support of this tube is
	\begin{equation}
		\begin{aligned}
			\Sigma_{\rm tube}^{(x)}(L,M,N)
			\coloneqq{}&
			(1+y^M)\Sigma_{xz}(L,N)
			+
			(1+z^N)\Sigma_{xy}(L,M)\\
			={}&
			\left(
			0,
			B
			\begin{pmatrix}
				xyzQ_{L,M,N}\\
				0
			\end{pmatrix}
			\right).
		\end{aligned}
		\label{eq:xcube_four_face_tube}
	\end{equation}
	The two \(xz\) faces and two \(xy\) faces have canceling corner charges,
	so the tube is gauge invariant with both \(yz\) ends open.
	Varying \(L,M,N\) gives generalized Wilson families with wireframe
	support in the \(Z\)-sector and open-ended tube support in the \(X\)-sector.
	
	\subsection{Checkerboard code}
	\label{subsec:checkerboard_gauge_supports}
	
	The checkerboard code is self-dual, so we study the \(X\)-sector.
	We use \(A=B=(\bar h,h)^T\), where \(h=1+u+v+w\) and
	\(\bar h=1+u^{-1}+v^{-1}+w^{-1}\). The induced gauge theory
	lives on the four-dimensional spacetime extension of the FCC lattice.
	Its interactions are specified by
	\begin{equation}
		(\Square_1^X)^\dagger
		=
		\begin{pmatrix}
			h&\bar h&0\\
			\bar\delta_t&0&\bar h\\
			0&\bar\delta_t&h
		\end{pmatrix}.
		\label{eq:checkerboard_interaction_incidence_map}
	\end{equation}

	Since \(\ker A=\ker B=0\), we take \(C_2^X=C_2^Z=0\).
	The gauge map \(\Square_2^X\) and its adjoint are
	\begin{equation}
		\Square_2^X
		=
		\begin{pmatrix}
			\delta_t\\
			\bar h\\
			h
		\end{pmatrix},
		\qquad
		(\Square_2^X)^\dagger
		=
		\begin{pmatrix}
			\bar\delta_t&h&\bar h
		\end{pmatrix}.
		\label{eq:checkerboard_gauge_maps}
	\end{equation}
	A gauge-invariant spin support
	\(\Sigma_X\in R_{\rm st}\oplus R_{\rm st}^2\) satisfies
	\(\mathfrak q_X[\Sigma_X]=(\Square_2^X)^\dagger\Sigma_X=0\).
	For a purely spatial support \((0,p)\) with \(p\in R_{\rm s}^2\),
	this condition becomes \(A^\dagger p=0\).

	Let \(e_1=(1,0)^T\) and \(e_2=(0,1)^T\) label the odd- and even-parity
	FCC orbits, represented by \((1,1,1)\) and \((0,0,0)\), respectively.
	Translation by one simple-cubic spacing exchanges the orbits.  In this
	basis of \(R_{\rm s}^2\), the one-step translation maps are
	\begin{equation}
		\begin{gathered}
			T_x=
			\begin{pmatrix}
				0&w^{-1}\\
				uv&0
			\end{pmatrix},
			\qquad
			T_y=
			\begin{pmatrix}
				0&v^{-1}\\
				uw&0
			\end{pmatrix},
			\qquad
			T_z=
			\begin{pmatrix}
				0&u^{-1}\\
				vw&0
			\end{pmatrix},
			\\[2pt]
			T_\alpha^2=\lambda_\alpha I_2,
			\qquad
			\lambda_x=uvw^{-1},\quad
			\lambda_y=uwv^{-1},\quad
			\lambda_z=vwu^{-1}.
		\end{gathered}
		\label{eq:checkerboard_one_step_translations}
	\end{equation}
	The three maps commute, and \(\lambda_\alpha\) translates the support by two
	lattice spacings along direction \(\alpha\).
	
	For distinct coordinate directions \(\alpha,\beta\) and positive integers
	\(\ell_\alpha,\ell_\beta\), define the spatial membrane support
	\begin{equation}
		p_{\alpha\beta}(\ell_\alpha,\ell_\beta)
		\coloneqq
		\sum_{a=0}^{\ell_\alpha-1}
		\sum_{b=0}^{\ell_\beta-1}
		T_\alpha^aT_\beta^b e_1 .
		\label{eq:checkerboard_rectangular_membranes}
	\end{equation}
	This support fills an \(\ell_\alpha\times\ell_\beta\) rectangle in the coordinate
	\(\alpha\beta\) plane, with its two components alternating under successive
	one-step translations.

	Let \(\gamma\) denote the direction transverse to the supporting
	\(\alpha\beta\) plane, so that
	\(\{\alpha,\beta,\gamma\}=\{x,y,z\}\).  The corresponding unit
	displacements are \(\hat\alpha,\hat\beta,\hat\gamma\).
	Applying the spatial charge map to
	the rectangular membrane produces four gauge charges at
	\begin{equation}
		\mathbf r_{ab}
		=
		a\ell_\alpha\hat\alpha
		+
		b\ell_\beta\hat\beta
		+
		\bigl[
			(a\ell_\alpha+b\ell_\beta)\bmod2
		\bigr]\hat\gamma,
		\qquad
		a,b\in\{0,1\},
		\label{eq:checkerboard_membrane_charge_positions}
	\end{equation}
	where the transverse offset keeps every charge representative on the even-parity
	FCC sublattice.  The four charges form a rectangle parallel to the supporting plane
	for even--even side lengths, a tilted rectangle when exactly one length is
	odd, and an irregular tetrahedron when both are odd.  The membrane itself
	remains planar.

	To construct a gauge-invariant order parameter, we use the even--even
	subfamily, whose charge pattern factorizes as
	\begin{equation}
		A^\dagger p_{\alpha\beta}(2L,2M)
		=
		(1+\lambda_\alpha^L)(1+\lambda_\beta^M).
		\label{eq:checkerboard_even_membrane_boundary}
	\end{equation}
	For \(L,M,N\geq1\), two pairs of parallel membranes define the purely spatial
	tube support
	\begin{equation}
		\Sigma_{\rm tube}(L,M,N)
		\coloneqq
		\begin{pmatrix}
			0\\[2pt]
			(1+\lambda_z^N)p_{xy}(2L,2M)
			+
			(1+\lambda_x^L)p_{yz}(2M,2N)
		\end{pmatrix}.
		\label{eq:checkerboard_tube_family}
	\end{equation}
	Equation~\eqref{eq:checkerboard_even_membrane_boundary} makes the two
	boundary patterns cancel, so \(\mathfrak q_X[\Sigma_{\rm tube}(L,M,N)]=0\).
	The four membranes form an open-ended rectangular tube; varying
	\(L,M,N\) gives a generalized Wilson family.

	\subsection{Haah's code}
	\label{subsec:haah_gauge_supports}

	Haah's cubic code is self-dual under spatial inversion and exchange of the
	two qubit species, so we study the \(X\)-sector.  In the conventions of
	Eq.~\eqref{eq:haah_code_stabilizer_map},
	\(A=(f,g)^T\) and \(B=(\bar g,\bar f)^T\), with
	\(f=1+x+y+z\) and \(g=1+xy+yz+zx\).
	The induced gauge theory lives on four-dimensional spacetime, with
	interaction map
	\begin{equation}
		(\Square_1^X)^\dagger
		=
		\begin{pmatrix}
			\bar f&\bar g&0\\
			\bar\delta_t&0&\bar g\\
			0&\bar\delta_t&\bar f
		\end{pmatrix}.
		\label{eq:haah_interaction_incidence_map}
	\end{equation}
	Since \(\ker A=\ker B=0\), we take \(C_2^X=C_2^Z=0\), giving
	\begin{equation}
		\Square_2^X
		=
		\begin{pmatrix}
			\delta_t\\ f\\ g
		\end{pmatrix},
		\qquad
		(\Square_2^X)^\dagger
		=
		\begin{pmatrix}
			\bar\delta_t&\bar f&\bar g
		\end{pmatrix}.
		\label{eq:haah_gauge_maps}
	\end{equation}
	Thus a support \(\Sigma_X=(u,v_1,v_2)^T\) is gauge invariant when
	\(\mathfrak q_X[\Sigma_X]=\bar\delta_t u+\bar f v_1+\bar g v_2=0\).
	Here \(u\) is the temporal component, while \(v_1,v_2\) are the two
	spatial spin components.

	To construct an explicit fractal family, let \(L=2^m\), \(m\geq1\), and
	choose a support on the first spatial spin component:
	\begin{equation}
		p_m\coloneqq
		\begin{pmatrix}
			\bar f^{L-1}\\ 0
		\end{pmatrix}.
		\label{eq:haah_fractal_creator}
	\end{equation}
	Over \(\mathbb F_2\), its nonzero component factorizes at successively
	doubled scales:
	\begin{equation}
		\bar f^{L-1}=\prod_{j=0}^{m-1}
		\bigl(1+\bar x^{2^j}+\bar y^{2^j}+\bar z^{2^j}\bigr).
		\label{eq:haah_fractal_factorization}
	\end{equation}
	This gives a sparse, self-similar pattern of \(4^m=L^2\) spins with
	spatial gauge-charge pattern
	\begin{equation}
		q_m\coloneqq A^\dagger p_m
		=\bar f^L=1+\bar x^L+\bar y^L+\bar z^L.
		\label{eq:haah_fractal_charge}
	\end{equation}
	The four charges lie at the tetrahedral corners
	\(0,-L\hat x,-L\hat y,-L\hat z\).

	For \(r\geq1\) measurement rounds, the suspension
	\eqref{eq:SI_wilson_suspension} gives
	\begin{equation}
		\Sigma_{\rm H}(m,r)\coloneqq
		\begin{pmatrix}
			t[r]_t q_m\\ (1+t^r)p_m
		\end{pmatrix}
		=(\Square_1^X)^\dagger
		\begin{pmatrix}
			t[r]_t p_m\\ 0
		\end{pmatrix},
		\qquad
		\mathfrak q_X[\Sigma_{\rm H}(m,r)]
		=(1+t^r)q_m+(1+t^r)q_m=0.
		\label{eq:haah_fractal_spacetime_support}
	\end{equation}
	This support consists of two identical spatial fractal patterns at rounds
	\(0,r\), joined by four temporal strands at the charge positions.
	The strands carry charges through time at fixed spatial positions; they
	do not describe spatial motion of isolated fractons.

	Let \(P_m=\operatorname{supp}\bar f^{L-1}\) and
	\(Q_m=\operatorname{supp}q_m\).  Writing
	\(\sigma_t,\sigma_1\in\{\pm1\}\) for the temporal and first spatial
	Ising-spin components, respectively, the
	corresponding generalized Wilson observable is
	\begin{equation}
		W_{\rm H}(m,r)
		=\prod_{\mathbf a\in P_m}
		\sigma_1(\mathbf a,0)\sigma_1(\mathbf a,r)
		\prod_{\mathbf b\in Q_m}\prod_{\tau=1}^{r}
		\sigma_t(\mathbf b,\tau).
		\label{eq:haah_fractal_wilson_observable}
	\end{equation}
	Increasing \(m\) and \(r\) gives a gauge-invariant family with sparse,
	non-manifold support at arbitrarily large spatial and temporal scales.
	As in the general construction,
	whether this family distinguishes phases is determined by its
	thermodynamic scaling.

	\emph{Remark (planar realization).}
	Although the support \(p_m\) above extends in all three spatial directions,
	a planar alternative is available.  The planar fractional moves in
	Eq.~(18) and Fig.~5(b) of Ref.~\cite{Song24} give, after relabeling
	the spatial axes,
	\begin{equation}
		\begin{gathered}
			h\coloneqq(x+y)f+g=1+x+x^2+y+xy+y^2,\\
			p_m^{xy}\coloneqq\bar h^{L-1}
			\begin{pmatrix}\bar x+\bar y\\1\end{pmatrix},
			\qquad
			A^\dagger p_m^{xy}=\bar h^L
			=1+\bar x^L+\bar x^{2L}+\bar y^L
			+\bar x^L\bar y^L+\bar y^{2L},
			\qquad L=2^m.
		\end{gathered}
		\label{eq:haah_planar_fractal_support}
	\end{equation}
	Both \(p_m^{xy}\) and its six-charge pattern \(\bar h^L\) involve
	only \(x,y\).  Their suspension by Eq.~\eqref{eq:SI_wilson_suspension}
	therefore gives two planar fractal patterns joined by six temporal
	strands, entirely within the \(xy\) plane times time.  Thus the support
	illustrated in Fig.~2(d) of the main text can be realized within a
	\((2+1)\)-dimensional subspace---two spatial directions and one temporal
	direction---of the four-dimensional gauge theory.

\end{document}